\documentclass[12pt,a4paper]{article}
\usepackage{amsmath,amssymb,mathrsfs,cite}
\usepackage{slashed}
\newcommand{\tr}{\operatorname{tr}}

\newcommand{\Det}{\operatorname{Det}}

\newcommand{\diag}{\operatorname{diag}}
\newcommand{\dif}{\mathrm{d}}

\newcommand{\pder}[2]{\frac{\partial #1}{\partial #2}}
\newcommand{\dder}[2]{\frac{\delta #1}{\delta #2}}
\newcommand{\iDelta}{\mathit{\Delta}}
\newcommand{\sgn}{\operatorname{sgn}}
\allowdisplaybreaks
\numberwithin{equation}{section}

\begin{document}
\begin{titlepage}
\begin{flushright}
UTHEP-791
\end{flushright}
\vspace{12mm}
\begin{center}
{\Large \bf
Quantisation of type IIB superstring theory and \\
the matrix model
}
\end{center}
\vspace{7mm}
\begin{center}
Yuhma Asano\footnote{
asano@het.ph.tsukuba.ac.jp
}

\par \vspace{7mm}
{\it
Institute of Pure and Applied Sciences, University of Tsukuba,\\
1-1-1 Tennodai, Tsukuba, Ibaraki 305-8571, Japan
}\\

{\it
Tomonaga Center for the History of the Universe, University of Tsukuba,\\
1-1-1 Tennodai, Tsukuba, Ibaraki 305-8571, Japan
}\\

\end{center}
\vspace{10mm}
\begin{abstract}\noindent
We discuss the path-integral quantisation of perturbative string theory
and show equivalence between the Polyakov-type, Schild-type and Nambu-Goto-type formulations of critical type II superstring theory
in the Minkowski and Euclidean signatures.
Remarkably, we also find that the Minkowskian path integral 
realises causality in the sense that a string does not propagate between points at space-like separation,
by giving careful consideration to the measure of the world-sheet metric.
We also discuss matrix regularisation of the path integral for type IIB perturbative superstring theory.
The obtained matrix models are the Euclidean IKKT matrix model and a modified Minkowskian IKKT model, 
depending on how the matrix regularisation is applied.

\end{abstract}
\setcounter{footnote}{0}
\end{titlepage}

\tableofcontents

\section{Introduction}
String theory is yet to be well-defined 
non-perturbatively in a satisfactory way.
To this end, the IKKT matrix model \cite{Ishibashi:1996xs}, 
also known as the IIB matrix model, 
was proposed as a non-perturbative formulation.
The matrix model is expected to describe type IIB superstring theory,
supported by various observations,
especially by the fact that
the action is obtained 
as matrix regularisation of type IIB perturbative superstring theory.
An important observation for reproducing our universe is that
the IKKT matrix model has potential to dynamically realise 4-dimensional spacetime at large $N$ \cite{Aoki:1998vn},
which is also supported by the spontaneous breaking of the target-space rotational symmetry $SO(10)$ to $SO(3)$ observed in numerical simulations\footnote{
See also 
Ref.~\cite{Anagnostopoulos:2022dak} for the recent development of Lorentzian IKKT simulations.
} \cite{Anagnostopoulos:2020xai,Kumar:2022giw}.

However,
explicit realisation of perturbative superstring theory from the matrix model
has not been well understood.
In particular, an important problem is 
how exactly gravitation is realised
and which matrix degrees of freedom correspond to gravitons,
which has been studied in terms of some interpretations, 
such as Hanada-Kawai-Kimura interpretation \cite{Hanada:2005vr}
and Weitzenb\"ock-connection interpretation \cite{Steinacker:2020xph,Steinacker:2019fcb},
in both of which the target-space metric is read off from the bosonic matrices.
On the other hand, there is no understanding of
how these interpretations are related to the original coordinate interpretation,
in which the bosonic matrices represent the coordinates of the target spacetime.

In addition, the IKKT matrix model has an issue typical of 0-dimensional theory.
In the canonical quantisation, a theory needs to have a temporal direction,
and the path integral formalism can be derived from time evolution of a state.
However, since there is no temporal direction in the IKKT matrix model
as a field theory,
the exact form of the path integral is undetermined 
without another principle---we do not really know 
whether the weight should be in the form of $e^{-S}$ or $e^{iS}$ 
and also whether the target-space metric should be Euclidean or Minkowski.

In order to resolve these problems, 
we revisit the matrix regularisation of perturbative string theory
as a first step.

The usual starting point of perturbative string theory is the Nambu-Goto-type action in the Minkowski signature.
However, since the Nambu-Goto type is not convenient 
for the path-integral quantisation 
as it involves the square root,
we mainly consider the Polyakov-type action
for perturbative computations.
In addition, since the Minkowski signature can be a source of obstacles,
we often naively Wick-rotate the theory to the Euclidean signature.
Thus, 
a string scattering amplitude is usually defined 
by Euclidean path integration with the Polyakov-type action \cite{Polyakov:1981rd},
which is now an established approach to perturbative string theory.

Although it is widely believed that the above procedures can be justified
for some reasons,
there has not been a rigorous proof of them but some partial ones.
As for the quantum equivalence between the Nambu-Goto-type and Polyakov-type formulations,
it was discussed for the Euclidean case in Ref.~\cite{Polyakov:1987ez}
and together with the Schild-type formulation \cite{Schild:1976vq} 
in Ref.~\cite{Yoneya:1997gs}.
On the other hand,
the equivalence between the Minkowskian path integral with the Polyakov-type action 
and its Euclidean version has not been shown
to the best of the author's knowledge.
In fact, a naive Wick rotation, $X^0=e^{-i\theta}X^{D}$,
where $D$ is the target-space dimension,
with rotation of the world-sheet vielbein
$e^0_{\;\, a}=e^{-i\theta}e^2_{\;\, a}$ 
(or more naively, $\sigma^0=e^{-i\theta}\sigma^2$) 
with $\theta$ varying from 0 to $\pi/2$, 
is not a valid contour deformation that equates the Minkowskian path integral to the Euclidean version by Cauchy's integral theorem
because the exponent of the path integrand is not bounded from above during the deformation.

In this paper, 
we provide a proof of the equivalence in terms of path integration
between the three types of formulation of critical type II superstring theory
in the Minkowski signature.
We also show the equivalence between the Minkowskian path integral 
and its Euclidean version.
Through discussion about the path-integration measure, 
we find an 
anti-fundamental string (anti-F-string) should be included in the path integration.
As a result, we see that causality is realised thanks to the existence of the anti-F-string.
Since the discussion of these equivalences provides 
a solid starting point for the path-integral formulation,
we use the matrix regularisation procedure to the path integral and
see which type of IKKT matrix model is derived.

The paper is organised as follows.
In section 2, 
we review how the Schild-type and Polyakov-type actions 
are classically derived from the Nambu-Goto-type action in the Green-Schwarz formalism.
In section 3, we show the equivalence of the three types of formulation
in terms of the path-integral quantisation
and see the realisation of causality.
Then, we discuss the gauge symmetry in the Schild-type action
in section 4, 
and 
we apply the matrix regularisation procedure to perturbative string theory 
to obtain the path integral for the matrix model
in section 5.
Finally, we summarise the paper and discuss some important issues of the matrix model in section 6.
In appendix,
we see 
the off-shell closure of a subalgebra of the local symmetry of the Schild-type action.

\section{Derivation of the Schild-type superstring action}
\label{sec:Schild-action}
Let us review how the Schild-type superstring action is derived.
We start with the Nambu-Goto-type action on the 10-dimensional flat spacetime
in the type II Green-Schwarz formalism \cite{Green:1983wt}:
\begin{align}
 S_{\rm NG}&=
 -\frac{1}{2\pi}
 \int \dif{^2\sigma} \bigg\{
 \sqrt{-h
 }
 -i\varepsilon^{ab}\partial_a X^\mu
 (\bar\theta^{1}\Gamma_\mu\partial_b\theta^1
 -\bar\theta^{2}\Gamma_\mu\partial_b\theta^2)
 \nonumber \\
 &\hspace{170pt}
 +\varepsilon^{ab}\bar\theta^{1}\Gamma^\mu\partial_a\theta^1
 \bar\theta^{2}\Gamma_\mu\partial_b\theta^2
 \bigg\}
 ,
 \label{NG-action}
\end{align}
where 
\begin{align*}
 \Pi_a^\mu:=
 \partial_a X^\mu-i
 (\bar\theta^{1}\Gamma^\mu\partial_a\theta^1
 +\bar\theta^{2}\Gamma^\mu\partial_a\theta^2)
 ,
\end{align*}
and $h:=\det h_{ab}$
with the induced metric $h_{ab}:=\Pi_a^\mu\Pi_{b\mu}$.
Here, $\mu,\nu$ are target-space indices running over the values $0,1,\cdots,9$ and
$a,b$ are world-sheet indices taking values of $0$ and $1$.
The target-space indices are contracted by the Minkowski metric $\eta_{\mu\nu}=\diag(-1,1,\cdots,1)$, and
$\varepsilon^{ab}$ is the totally antisymmetric tensor with $\varepsilon^{01}=1$.
$\Gamma_\mu$'s are the gamma matrices for $Spin(9,1)$. 
$\theta^A$ ($A=1,2$) are two Majorana-Weyl spinors\footnote{
For simplicity, we take a representation in which the charge conjugation matrix
can be regarded as proportional to an identity matrix for Weyl spinors:
$\bar\theta^A=\theta^{AT}$ and $-\theta^{AT}$ 
for left-handed and
right-handed fermions, respectively.
}.
Note $\alpha'$ is absorbed to $X^\mu$: it can be restored 
via $X^\mu\to \frac{1}{\sqrt{\alpha'}}X^\mu$.
Hereafter, we use the following shorthand notations:
\begin{align}
 \slashed\Pi_{a}:=\Pi_a^\mu\Gamma_{\mu},
 \qquad
 \slashed\Pi_{ab}:=\Pi_a^\mu\Pi_b^\nu\Gamma_{\mu\nu},
\end{align}
where $\slashed\Pi_{a}$ is regarded as the induced gamma matrices.
We denote the inverse of $h_{ab}$ by $h^{ab}$.

The theory has 
bosonic and fermionic gauge symmetries:
the reparametrisation invariance ($\sigma^a\to\sigma^a+\delta\sigma^a$)
\begin{align}
 \delta \theta^A=-\delta\sigma^a\partial_a\theta^A,
 \qquad
 \delta X^\mu=-\delta\sigma^a\partial_aX^\mu
\end{align}
and
the $\kappa$ symmetry
\begin{align}
 \delta\theta^1=(\mathbf{1}+\tilde\Gamma)\kappa^1,
 \qquad
 \delta\theta^2=(\mathbf{1}-\tilde\Gamma)\kappa^2,
 \qquad
 \delta X^\mu=-i
 (\delta\bar\theta^{1}\Gamma^\mu\theta^1
 +\delta\bar\theta^{2}\Gamma^\mu\theta^2)
 ,
\end{align}
where
\begin{align}
 \tilde\Gamma=\frac{\varepsilon^{ab}}{2\sqrt{-h}}\slashed\Pi_{ab}
 ,
\end{align}
which satisfies $\tilde\Gamma^2=\mathbf{1}$.

One can derive the Schild-type action 
and the Polyakov-type action
from the Hamiltonian formalism 
with constraints
obtained by the Legendre transformation of the Nambu-Goto-type action \eqref{NG-action}:
\begin{align}
 \mathcal{H}
 &=\Lambda^0
 \chi_{b0}
 +\Lambda^1
 \chi_{b1}
 +\chi_f^{1\, T}\Lambda_f^1
 +\chi_f^{2\, T}\Lambda_f^2
 ,
 \label{string-Hamiltonian}
\end{align}
where $\Lambda^a$ and $\Lambda_f^A$ are the Lagrange multipliers for 
the bosonic and fermionic constraints: 
\begin{align}
 &\chi_{b0}
 =\pi\left\{
 \left(
 P_\mu
 -\frac{i}{2\pi}(\bar\theta^{1}\Gamma_{\mu}\partial_1\theta^1-\bar\theta^{2}\Gamma_{\mu}\partial_1\theta^2)
 \right) ^2
 +\frac{1}{(2\pi)^2}\Pi_{1}^\mu\Pi_{1\mu}
 \right\}
 \approx 0,
 \label{chi_b0} \\
 &\chi_{b1}
 =P_\mu\partial_1 X^\mu
 + \pi^{1\, T}\partial_1\theta^1 + \pi^{2\, T}\partial_1\theta^2
 \approx 0,
 \label{chi_b1} \\
 &\chi_f^{1\, T}
 =\pi^{1\, T}
 +i\left(
 P_\mu
 +\frac{\partial_1X_\mu}{2\pi}
 -\frac{i}{2\pi}\bar\theta^{1}\Gamma_{\mu}\partial_1\theta^1
 \right) (\bar\theta^{1}\Gamma^\mu)
 \approx 0,
 \label{chi_f1} \\
 &\chi_f^{2\, T}
 =\pi^{2\, T}
 +i\left(
 P_\mu
 -\frac{\partial_1X_\mu}{2\pi}
 +\frac{i}{2\pi}\bar\theta^{2}\Gamma_{\mu}\partial_1\theta^2
 \right) (\bar\theta^{2}\Gamma^\mu)
 \approx 0
 \label{chi_f2}
 .
\end{align}
Here,
$P_\mu$ and $\pi^A$ are the conjugate momenta for $X^\mu$ and $\theta^A$,
respectively.
Note that the fermionic constraints 
projected by $\frac{1}{2}(\mathbf{1}+(-1)^{A+1}\tilde\Gamma)$
are first class,
which corresponds to the fermionic gauge symmetry,
and the other half are second class.

Then, again by the Legendre transformation, 
integrating out $\pi^A$, $\Lambda_f^A$ and $P_\mu$,
one obtains the Polyakov-type action:
\begin{align}
 S_{\rm P}
 &=-\frac{1}{2\pi}\int \dif{^2\sigma} \, \bigg\{
 \frac{1}{2}\lambda^{ab}h_{ab}
 -i\varepsilon^{ab}\partial_a X^\mu
 (\bar\theta^{1}\Gamma_\mu\partial_b\theta^1
 -\bar\theta^{2}\Gamma_\mu\partial_b\theta^2)
 +\varepsilon^{ab}\bar\theta^{1}\Gamma^\mu\partial_a\theta^1
 \bar\theta^{2}\Gamma_\mu\partial_b\theta^2
 \bigg\}
 ,
 \label{Polyakov-action}
\end{align}
where $\lambda^{00}=-\frac{1}{\Lambda^0}$ and
$\lambda^{01}=\lambda^{10}=\frac{\Lambda^1}{\Lambda^0}$,
$\lambda^{11}=((\lambda^{01})^2-1)/\lambda^{00}$,
which are written by the world-sheet metric $g_{ab}$ as $\lambda^{ab}=\sqrt{-g}g^{ab}$.
Its $\kappa$ symmetry is 
\begin{align}
 &\delta\theta^1
 =\left(
 \mathbf{1}
 +\frac{\lambda^{cd}h_{cd}}{4(-h)}\varepsilon^{ab}\slashed\Pi_{ab}
 \right) \kappa^1
 ,
 \qquad
 \delta\theta^2
 =\left(
 \mathbf{1}
 -\frac{\lambda^{cd}h_{cd}}{4(-h)}\varepsilon^{ab}\slashed\Pi_{ab}
 \right) \kappa^2
 ,
 \nonumber \\
 &\delta X^\mu=-i(\delta\bar\theta^1\Gamma^\mu\theta^1+\delta\bar\theta^2\Gamma^\mu\theta^2)
 ,
 \quad
 \delta\lambda^{ab}
 =8i
 (\bar\kappa^1_c\mathcal{P}_+^{ac}\mathcal{P}_+^{bd}\partial_d\theta^1
 +\bar\kappa^2_c\mathcal{P}_-^{ac}\mathcal{P}_-^{bd}\partial_d\theta^2)
 ,
 \label{kappa-symmetry_P}
\end{align}
where 
$\mathcal{P}_\pm^{ab}:=\frac{1}{2}(\lambda^{ab}\pm\varepsilon^{ab})$
and 
$\kappa^A_a:=\frac{1}{-h}\varepsilon_{ab}\lambda^{bc}h_{cd}\varepsilon^{de}\slashed\Pi_e\kappa^A$.
Note $\delta\theta^A=\mathcal{P}_{(-)^{A+1}}^{ab}\slashed\Pi_a\kappa_b^A$.

Furthermore, integration-out over $\Lambda^1$ leads to
the Schild-type action:
\begin{align}
 S_{\rm Schild}
 &=-\frac{1}{2\pi}\int \dif{^2\sigma} \, \bigg\{
 \frac{1}{2}\left(
 \frac{-h}{e_g}
 +e_g
 \right)
 -i\varepsilon^{ab}\partial_a X^\mu
 (\bar\theta^{1}\Gamma_\mu\partial_b\theta^1
 -\bar\theta^{2}\Gamma_\mu\partial_b\theta^2)
 \nonumber \\
 &\hspace{170pt}
 +\varepsilon^{ab}\bar\theta^{1}\Gamma^\mu\partial_a\theta^1
 \bar\theta^{2}\Gamma_\mu\partial_b\theta^2
 \bigg\}
 ,
 \label{Schild-action}
\end{align}
changing the remaining Lagrange multiplier 
to $e_g$ by
\begin{align}
 e_g=h_{11}\Lambda^0.
\end{align}
Therefore, through Dirac's method of quantisation,
the Schild-type action should be quantum-mechanically equivalent 
to the Nambu-Goto type.
As is obvious from the above derivation,
one can easily check the equivalence at the classical level:
integrating out $e_g$
using the equation of motion 
\begin{align}
 e_g^2=-h
 ,
\end{align}
one reproduces the Nambu-Goto-type action \eqref{NG-action} and hence its equations of motion.

Before closing the review section,
let us point out that
the Schild-type theory contains exactly the same constraints as 
the Nambu-Goto-type theory.
The expressions of the momenta derived from the Schild-type action
immediately lead us to Eq.~\eqref{chi_b1}, \eqref{chi_f1} and \eqref{chi_f2}.
Then the Poisson bracket of
the Hamiltonian density $\mathcal{H}_{\rm Schild}=\frac{e_g}{h_{11}}\chi_{b0}$
and Eq.~\eqref{chi_b1} provides Eq.~\eqref{chi_b0}.
Therefore,
the number of the gauge degrees of freedom in the Schild-type action is 
2 bosonic and 16 fermionic degrees of freedom.

\section{Path integral for perturbative string theory}
\label{sec:path-integral}
The derivation in the previous section does not answer
how exactly the Schild-type formulation is related to the Nambu-Goto type
in the path-integral formalism.
In this section,
let us discuss the path-integral definition of perturbative superstring theory
and see the relationship between 
the Nambu-Goto-type, Polyakov-type and Schild-type formulations.
The discussion in this section is applicable to type IIA superstring theory as well.

Since the perturbation theory formulated by the Polyakov-type action is well-established,
we start with its path integral with a complex-phase weight factor whose action is in the Minkowski signature,
\begin{align}
 Z
 &=\int 
 \mathcal{D}X\, \mathcal{D}\theta\,
 Z_{\rm b}[X,\theta]\,
 \exp\bigg[
 -\frac{i}{2\pi}\int \dif{^2\sigma} \, \bigg\{
 -i\varepsilon^{ab}\partial_a X^\mu
 (\bar\theta^{1}\Gamma_\mu\partial_b\theta^1
 -\bar\theta^{2}\Gamma_\mu\partial_b\theta^2)
 \nonumber \\
 &\hspace{250pt}
 +\varepsilon^{ab}\bar\theta^{1}\Gamma^\mu\partial_a\theta^1
 \bar\theta^{2}\Gamma_\mu\partial_b\theta^2
 \bigg\}
 \bigg]
 ,
 \label{Polyakov-path-int}
\end{align}
where 
\begin{align}
 Z_{\rm b}[X,\theta]
 &=\int 
 \mathcal{D}g\,
 \exp\bigg[
 -\frac{i}{2\pi}\int \dif{^2\sigma}\,
 \frac{1}{2}\lambda^{ab}h_{ab}
 \bigg]
 .
 \label{partition-fn_b}
\end{align}
Note that the world-sheet metric is written by the Lagrange multipliers $\Lambda^a$ and the conformal factor $\phi$
as $g_{ab}=e^\phi(\lambda^{-1})_{ab}$.
The measure of integration should be invariant under a reparametrisation transformation\footnote{The measure cannot be invariant under a $\kappa$ transformation,
but instead, it should be transformed so that 
its transformation is compatible with the algebra of the gauge symmetry
\cite{DeWitt:1984sjp}.
}.
This Minkowskian path integral is related, 
by Wick rotation, to the Euclidean version, 
which is usually used in computation of string scattering amplitudes;
they are indeed equivalent as discussed later.
One can also obtain this path integral
from the Minkowskian path integral of
the Hamiltonian with the constraints \eqref{string-Hamiltonian},
by integrating out $\pi^A$, $\Lambda_f^A$, and then $P_\mu$.
From now on, let us concentrate on the ``bosonic'' part $Z_{\rm b}[X,\theta]$ 
since the three types of formulation share the rest ``fermionic'' terms
in the action in Eq.~\eqref{Polyakov-path-int}.

Note that the path integral is not well-defined at this point without regulators
with infinitesimal cut-off parameters
because the Minkowskian path integral is not absolutely convergent.
However, regulators for integration over $X^\mu$ and $\Lambda^a$ naturally enter into the action
as the so-called ``$i\epsilon$ terms,''\footnote{
The $i\epsilon$ prescription in string theory was discussed 
in Ref.~\cite{Berera:1992tm,Witten:2013pra}
for scattering amplitudes of Euclidean theory 
but with the target space Wick-rotated to the Minkowski signature.
The prescription in this paper differs in that
$i\epsilon$ terms are implemented in the path-integral weight
before computation of scattering amplitudes,
which should be somehow related to their prescription.
}
like standard field theories,
when the path integral is for an expectation value in a vacuum.
In perturbative string theory, the S-matrix is usually defined by a path integral with insertions of vertex operators each of which corresponds to a single-string state.
In the state--operator correspondence, 
each vertex operator contains a factor 1, which is mapped to the single-string ground state;
therefore 
the S-matrix path integral 
should be implicitly assumed to
contain the ground-state wave functions.
If we fix the gauge symmetry of the Polyakov-type action 
by the light-cone gauge $X^+=\sigma^0$
with $g_{ab}=\eta_{ab}$,
the bosonic part of the ground-state wave function of a single closed string is proportional to
$\exp[-\sum_{n=1}^\infty \sum_{i=1}^{D-2} n X_{-n}^i X_{n}^i]$, 
where $X^i_n$ 
is a mode of $X^i$ of momentum $n$.
Thus, in and out ground-state wave functions effectively give
$i\epsilon \Pi_{1}^i\Pi_{1}^i$
to the Lagrangian density.
Actually, for the sake of regulating the divergences of the action as $\Lambda^0\to 0$ and $\infty$
in a gauge-invariant way\footnote{
Other than the diffeomorphism,
the Polyakov-type action and its integration measure
has another strange, but trivial gauge symmetry:
$\delta(\sqrt{-g}g^{ab})=\alpha\,\varepsilon^{c(a}h_{cd}\sqrt{-g}g^{b)d}$
with the other fields unchanged.
Then the summation of the diffeomorphism with $\delta\sigma^a$ and 
this gauge symmetry with $\alpha=\frac{2}{h_{11}}\partial_1\delta\sigma^0$
is the remaining bosonic gauge symmetry in the Schild-type action.
The regulator respects this combined gauge symmetry.
}, 
we should reinterpret
the regulator 
as a shift of $\Lambda^0$ 
to $\Lambda^0-i\epsilon|\Lambda^0|$
that appears in the terms coupled to $\Pi_a^i\Pi_b^i$,
providing a regulator with an additional term: 
\begin{equation}
 i\epsilon \sum_{i=1}^{D-1}
 \left(|\Lambda^0|\,\Pi_1^i\Pi_1^i+\frac{1}{|\Lambda^0|}(\Pi_0^i-\Lambda^1\Pi_1^i)^2\right),
 \label{cutoff-regulator}
\end{equation}
which is still positive.
This regulator is not only for $X^i$ but also for $\Lambda^0$ and $\Lambda^1$,
which should originally come from the regulators 
implicitly contained in the imposition of the constraints:
$\delta(\chi)
=\lim_{\epsilon\to 0}\int_{-\infty}^{\infty}\frac{\dif{\Lambda}}{2\pi}\,e^{i(\Lambda\chi+i\epsilon|\Lambda|)}$.

Apparently, this regulator
seems to be incomplete and also to break Lorentz symmetry
since it does not regularise the path integration over the temporal field $X^0$.
However, we know 
that excitation of such unphysical degrees of freedom 
does not contribute to the amplitude
in critical superstring theory. 
Therefore, the path integral should be independent of the regularisation for such degrees of freedom,
and the integration is expected to be well-defined.

To integrate out $\Lambda^a$ 
with the regulator taken into account, 
we first decompose the measure $\mathcal{D}g$,
which is induced by the reparametrisation-invariant norm
\begin{align}
 \frac{1}{2}\int \dif^2\sigma \sqrt{-g}\, 
 g^{da} \iDelta g_{ab}\, g^{bc} \iDelta g_{cd}
 =\int \dif^2\sigma\, e^{\phi}
 \left(
 \frac{(\iDelta\Lambda^0)^2-(\iDelta\Lambda^1)^2}{(\Lambda^0)^2}
 +\iDelta\phi^2
 \right)
 ,
\end{align}
for infinitesimal $\iDelta\Lambda^{a}$ and $\iDelta\phi$,
into
\begin{align}
 \mathcal{D}g
 =\mathcal{D}\phi\, \mathcal{D}\Lambda^0\, \mathcal{D}\Lambda^1
 \Det\left[ \frac{e^\phi}{(\Lambda^0)^2} \right]
 ,
\end{align}
where 
$\mathcal{D}\phi$ and $\mathcal{D}\Lambda^a$ are defined by
\begin{align}
 1
 =\int\mathcal{D}\phi\, e^{-\frac{\pi}{4}\int\dif{^2\sigma}\sqrt{-g}\, \phi^2} 
 =\int\mathcal{D}\Lambda^0\, e^{-\pi\int\dif{^2\sigma}\,(\Lambda^0)^2}
 =\int\mathcal{D}\Lambda^1\, e^{-\pi\int\dif{^2\sigma}\,(\Lambda^1)^2}
 .
\end{align}
The intervals of integration over $\Lambda^a$ and $\phi$ are all $(-\infty,\infty)$ at each point $\sigma$,
which agrees with the origin of $\Lambda^a$ being the corresponding constraints.
We will provide further explanation of the interval of $\Lambda^0$ later in this section.
As we are considering critical string theory,
the integration over the conformal factor $\phi$ does not produce the conformal anomaly\footnote{
A difference of the conformal anomaly 
between the Polyakov type and the Nambu-Goto type
was discussed for non-critical bosonic string theory in the Euclidean signature 
in Ref.~\cite{Makeenko:2021hcm,Makeenko:2023fzt}.
},
and it trivially becomes proportional to $[\prod_{\sigma}\iDelta\Sigma]$
since $e^\phi$ transforms in the same manner 
under the diffeomorphism
as the inverse of the infinitesimal world-sheet area element $\iDelta\Sigma$.
Thus the measure $\mathcal{D}g$ after the integration over $\phi$ can be written as
\begin{align}
 \mathcal{D}\lambda:=
 \lim_{\iDelta\Sigma\to 0}
 \prod_{\sigma}\frac{\dif\Lambda^0 \dif\Lambda^1}{(\Lambda^0(\sigma))^2}
 .
\end{align}
Though there could possibly be a subtlety in regard to the modular parameters, 
which are all included in $\Lambda^a$,
we assume it does not affect our argument of the relationship among the three formulations.

We then integrate out $\Lambda^1$ 
to connect the Polyakov-type and Schild-type formulations.
The path integral \eqref{partition-fn_b} becomes
\begin{align}
 Z_{\rm b}[X,\theta]
 &=\int 
 \mathcal{D}\lambda\,
 \exp\bigg[
 -\frac{i}{2\pi}\int \dif{^2\sigma}\,
 \left(
 \frac{1}{2}\lambda^{ab}h_{ab}
 -i\epsilon|\Lambda^0|
 -\frac{i\tilde\epsilon}{|\Lambda^0|}(\Lambda^1-c)^2
 \right)
 \bigg]
 \nonumber \\
 &=\int
 \mathcal{D}e_g\,
 \exp\bigg[
 -\frac{i}{2\pi}\int \dif{^2\sigma}\,
 \frac{1}{2}\left(
 \frac{-h}{e_g}
 +e_g
 -i\epsilon'|e_g|
 -\frac{i\tilde\epsilon'}{|e_g|}
 \right)
 \bigg]
 ,
 \label{partition-fn_b_PtoS}
\end{align}
with
\begin{align}
 \mathcal{D}e_g
 :=\lim_{\iDelta\Sigma\to 0}
 \prod_{\sigma}\frac{2\pi\,\dif e_g}{-(ie_g)^\frac{3}{2}\sqrt{\iDelta\Sigma}}
 ,\label{measure_eg}
\end{align}
where we change variables by $e_g=h_{11}\Lambda^0$ and
introduce the regulators \eqref{cutoff-regulator} 
with $\epsilon$, $\tilde\epsilon$ and $c$
ignoring the trivial $X^i$-dependence in them,
and redefining them by $\epsilon'$ and $\tilde\epsilon'$ on the last line.
The equation shows that the Polyakov-type and Schild-type formulations are
equivalent as long as the measures are properly set.

Finally, integrating out $e_g$ gives the Nambu-Goto-type formulation.
Using
\begin{align}
 \int_0^\infty \frac{\dif x}{-(ix)^{\frac{3}{2}}}\,
 \exp\left[
 -i\alpha\left(
 \frac{\beta}{x}+x
 \right)
 \right]
 =\frac{\sqrt{\pi}}{\alpha^{\frac{1}{2}}\beta^{\frac{1}{2}}}e^{-2i\alpha\beta^{\frac{1}{2}}}
 ,
 \label{eg-int-formula}
\end{align}
which holds 
for $\operatorname{Im}(\alpha)\leq 0$ and $\operatorname{Im}(\alpha\beta)\leq 0$,
the path integral~\eqref{partition-fn_b_PtoS} becomes
\begin{align}
 Z_{\rm b}[X,\theta]
 &=\prod_{\sigma}
 \bigg(
 \frac{4\pi^2}{\iDelta\Sigma\sqrt{-h-i\epsilon''}}
 \exp\bigg[
 -\frac{i}{2\pi}\iDelta\Sigma\,
 \sqrt{-h-i\epsilon''}
 \bigg]
 \nonumber \\
 &\hspace{60pt}
 +
 \frac{4\pi^2}{\iDelta\Sigma\sqrt{-h+i\epsilon''}}
 \exp\bigg[
 \frac{i}{2\pi}\iDelta\Sigma\,
 \sqrt{-h+i\epsilon''}
 \bigg]
 \bigg)
 \nonumber \\
 &=\left[ \prod_{\sigma}\sum_{s(\sigma)=\pm 1} \right]
 \!\!\!
 \left[ \prod_{\sigma}\frac{4\pi^2}{\iDelta\Sigma\sqrt{-h-i\epsilon''s(\sigma)}} \right]
 \!
 \exp\bigg[
 -\frac{i}{2\pi}\int\dif{^2\sigma}\,
 s(\sigma)\sqrt{-h-i\epsilon''s(\sigma)}
 \bigg]
 ,
 \label{partition-fn_b_NG}
\end{align}
with the infinitesimal regulators redefined by $\epsilon''$.
Remarkably,
the Nambu-Goto-type formulation
equivalent to the Polyakov type and the Schild type
has an ensemble in which the Lagrangian with the opposite overall sign
equally contributes
so that the weight involves $s(\sigma)=\pm 1$.
Moreover, the ensemble does not contain real configurations with $h>0$
since the terms on the first and second lines on the right-hand side
of Eq.~\eqref{partition-fn_b_NG} cancel as $\epsilon''\to 0$.
Therefore, configurations in which a string propagates between points at space-like separation are prohibited.
The cancellation is a natural consequence 
because the contour of the integration over $e_g$ in Eq.~\eqref{partition-fn_b_PtoS}
for $h>0$
can be deformed 
by $e_g=e^{-i\sgn(\tilde e_g)\theta}\tilde e_g$ 
with $\tilde e_g$ being a real variable running from $-\infty$ to $+\infty$
and with $\theta$ varying from $0$ to $\pi/2$,
which results in manifest cancellation between the integrals over $\tilde e_g>0$ and $\tilde e_g<0$.

One can interpret this cancellation as realisation of causality
by existence of strings corresponding to anti-particles.
First of all,
the two regions of integration, $\Lambda^0>0$ and $\Lambda^0<0$, 
are disconnected in terms of the world-sheet diffeomorphism
since $g_{ab}$ has divergent elements as $\Lambda^0\to 0$ and $\infty$.
This could propose another possibility of 
the path-integration measure---the integration over $\Lambda^0$ is only on $(0,+\infty)$
despite the origin of $\Lambda^0$,
which could be more or less a natural choice as well
because the coefficient of $P_i^2$ in the Hamiltonian \eqref{string-Hamiltonian} 
is $\pi\Lambda^0$.
We refer to such a configuration as ``positive-energy'' in this paper
because one can regard the $P_i^2$ term as a kinetic term in the right sign
although the Hamiltonian is not necessarily positive---actually,
the Hamiltonian should be classically zero.
If one chooses such a measure, 
then the interval of integration over $e_g$
on the right-hand side of Eq.~\eqref{partition-fn_b_PtoS}
is $(0,+\infty)$ or $(-\infty,0)$ depending on the sign of $h_{11}$;
alternatively, one can fix the interval to $(0,+\infty)$
by replacing every $e_g$ that appears in the integrand by $\sgn(h_{11}) e_g$.
After the integration over $e_g$, one obtains
\begin{align*}
 \left[ \prod_{\sigma}\frac{4\pi^2}{\iDelta\Sigma\sqrt{-h-i\epsilon''\sgn(h_{11})}} \right]
 \exp\bigg[
 -\frac{i}{2\pi}\int\dif{^2\sigma}\,
 \sgn(h_{11})\sqrt{-h-i\epsilon''\sgn(h_{11})}
 \bigg]
 .
\end{align*}
Since
$h_{11}$ should be classically positive for a propagating free string, 
this path integral corresponds to the contribution with $s(\sigma)=1$ in Eq.~\eqref{partition-fn_b_NG}.
In the same way, 
the path integral over $\Lambda^0$ on $(-\infty,0)$,
the ``negative-energy'' contribution,
corresponds to the one with $s(\sigma)=-1$,
which should be interpreted as an anti-F-string
since $s(\sigma)$ flips only the sign of the area term $\sqrt{-h}$.
Therefore, the contributions from $s(\sigma)=1$ and $-1$ together in Eq.~\eqref{partition-fn_b_NG}
take account of a ``positive-energy'' F-string and a ``negative-energy'' anti-F-string,
and importantly,
they both are needed for the causality, 
like standard quantum field theory.
It is also expected that
the existence of anti-F-strings in the path integral
should reproduce
the energy spectrum in quantum field theory\footnote{
See Ref.~\cite{Gavrilov:2000cn} for related discussion 
for a relativistic point particle
in the canonical quantisation.
}.

Another additional remark is that
the Minkowskian path integral \eqref{Polyakov-path-int}
is equivalent to its Euclidean version 
in the case of bosonic string theory
and type II superstring theory, by Wick rotation,
as shown in section \ref{sec:equiv-to-Euc_gauge-fixed}.
One can also check this by analytically continuing 
the Minkowskian path integral with $Z_{\rm b}$ in Eq.~\eqref{partition-fn_b_NG} 
by $\sigma^0=e^{-is\theta'}\sigma^2\to-is\sigma^2$
and then deforming the contour by $X^0=e^{-is\theta}X^{{\rm (E)}D}$ 
with $\theta\to\pi/2$,
assuming that the path integral is independent of the direction in which $\iDelta\Sigma$ approaches $0$.
The equivalence to the Euclidean path integral
indicates that the anti-F-string contribution is somehow included
in the computations of Polyakov's Euclidean path-integral.

\section{Gauge fixing}
\label{sec:gauge-fixing}
The Schild-type action is invariant under the gauge transformations
corresponding to those in the Nambu-Goto-type action,
as a matter of course,
since the two actions share the constraints of the system.
In this section, we study the gauge symmetry of the Schild-type action
of type IIB Green-Schwarz superstring theory,
and fix the gauge symmetry to obtain the path integral 
suitable for the connection to the IKKT matrix model.
We also show the equivalence of the Minkowskian superstring theory 
to its Euclidean version
for the sake of matrix regularisation after Wick rotation, 
which will be discussed in section \ref{sec:mat-reg_after-W}.

\subsection{Local fermionic symmetry for the Schild type}
Let us first change the type IIB fermionic variables,
which are both left-handed, by
\begin{align}
 \varphi=\frac{1}{2}(\theta^1+i\theta^2)
 ,
 \qquad
 \psi=\frac{1}{2}(\theta^1-i\theta^2)
 ,
\end{align}
with the path-integral contour for $\theta^2$ rotated by $\theta^2=i\tilde\theta^2$
with Majorana-Weyl $\tilde\theta^2$.
The Schild-type action~\eqref{Schild-action} then becomes
\begin{align}
 S_{\rm Schild}[X,\psi,\varphi,e_g]
 &=\frac{1}{2\pi}\int \dif{^2\sigma} \, e_g\bigg[
 \frac{h}{2e_g^2}
 -\frac{1}{2}
 +\frac{2i\varepsilon^{ab}}{e_g}
 (\psi^T\Gamma_\mu
 \partial_aX^\mu \partial_b\psi
 +\varphi^T\Gamma_\mu
 \partial_aX^\mu \partial_b\varphi
 )
 \nonumber \\
 &
 \hspace{50pt}
 -\frac{2\varepsilon^{ab}}{e_g}
 (\psi^T\Gamma^\mu\partial_a\psi
 +\varphi^T\Gamma^\mu\partial_a\varphi)
 (\psi^T\Gamma_\mu\partial_b\varphi
 +\varphi^T\Gamma_\mu\partial_b\psi)
 \bigg]
 ,
\end{align}
with $\Pi_a^\mu=\partial_aX^\mu-2i(\psi^T\Gamma^\mu\partial_a\varphi+\varphi^T\Gamma^\mu\partial_a\psi)$.
Just like the Nambu-Goto-type theory,
it has the reparametrisation symmetry
\begin{align}
 \delta \theta^A=-\delta\sigma^a\partial_a\theta^A,
 \qquad
 \delta X^\mu=-\delta\sigma^a\partial_aX^\mu,
 \qquad
 \delta e_g=-\partial_a(\delta\sigma^a e_g)
 .
 \label{reparam-gauge-symmetry}
\end{align}

The local fermionic symmetry of the Schild-type action is now
\begin{align}
 \delta X^\mu
 &=-i
 (\delta\theta^{1T}\Gamma^\mu\theta^1
 +\delta\theta^{2T}\Gamma^\mu\theta^2)
 =-2i(\delta\psi^{T}\Gamma^\mu\varphi
 +\delta\varphi^{T}\Gamma^\mu\psi)
 ,
 \nonumber \\
 \delta e_g
 &=\frac{4ie_g^2}{e_g^2+h}\sum_{A=1}^{2}\left( \frac{-h}{e_g}h^{ab}+(-1)^{A+1}\varepsilon^{ab}\right)
 \delta\theta^{AT}\slashed\Pi_{a}\partial_{b}\theta^A
 \nonumber \\
 &=\frac{8ie_g^2}{e_g^2+h}\bigg[
 \delta\varphi^{T}\slashed\Pi_{b}
 \left( \frac{-h}{e_g}h^{ab}\partial_{a}\psi-\varepsilon^{ab}\partial_a\varphi\right)
 +\delta\psi^{T}\slashed\Pi_{b}
 \left( \frac{-h}{e_g}h^{ab}\partial_{a}\varphi-\varepsilon^{ab}\partial_a\psi\right)
 \bigg]
 \label{kappa-symmetry_Schild}
 ,
\end{align}
for any infinitesimal 
$\delta\varphi=\frac{1}{2}(\delta\theta^1+i\delta\theta^2)$ 
and
$\delta\psi=\frac{1}{2}(\delta\theta^1-i\delta\theta^2)$ 
without any projection operator like $\frac{1}{2}(\mathbf{1}\pm\tilde\Gamma)$.
By restricting the transformation 
to
\begin{equation*}
 \delta\theta^1=\left( \mathbf{1}+\frac{\varepsilon^{ab}}{2e_g}\slashed\Pi_{ab} \right) \kappa^1,
 \qquad
 \delta\theta^2=\left( \mathbf{1}-\frac{\varepsilon^{ab}}{2e_g}\slashed\Pi_{ab} \right) \kappa^2,
\end{equation*}
or equivalently
\begin{align}
 \delta\varphi=\kappa^{\varphi}+\frac{\varepsilon^{ab}}{2e_g}\slashed\Pi_{ab} \kappa^\psi,
 \qquad
 \delta\psi=\kappa^{\psi}+\frac{\varepsilon^{ab}}{2e_g}\slashed\Pi_{ab} \kappa^{\varphi},
 \label{kappa-gauge-symmetry}
\end{align}
where $\kappa^{\varphi}=\frac{1}{2}(\kappa^1+i\kappa^2)$ and
$\kappa^{\psi}=\frac{1}{2}(\kappa^1-i\kappa^2)$,
one obtains the $\kappa$ symmetry equivalent to the Nambu-Goto type on-shell.
In this case, the transformation of $e_g$ becomes
\begin{align}
 &\delta e_g
 =4i\varepsilon^{ab}(\kappa^{1T}\slashed\Pi_{a}\partial_{b}\theta^1-\kappa^{2T}\slashed\Pi_{a}\partial_{b}\theta^2)
 =8i\varepsilon^{ab}(\kappa^{\varphi T}\slashed\Pi_{a}\partial_{b}\varphi+\kappa^{\psi T}\slashed\Pi_{a}\partial_{b}\psi)
 ,
\end{align}
without the on-shell divergent factor, $1/(e_g^2+h)$.

In addition, there is a trivial local bosonic symmetry:
\begin{align}
 \delta X^\mu=0,
 \qquad
 \delta\psi=0,
 \qquad
 \delta\varphi=0,
 \qquad
 \delta e_g=\frac{e_g^2}{e_g^2+h}\partial_a(e_g\mu^a)
 ,
 \label{mu-symmetry}
\end{align}
for infinitesimal $\mu^a$.
The finite form of the transformation \eqref{mu-symmetry} is
actually not divergent on-shell.
It is written as
\begin{align}
 \delta e_g
 =\frac{e_g^2+h}{2e_g}
 \Bigg(
 \sqrt{1
 +\frac{2e_g(e_g^2-h)}{(e_g^2+h)^2}\partial_a(e_g\mu^a)
 +\frac{e_g^2}{(e_g^2+h)^2}(\partial_a(e_g\mu^a))^2}
 -1+\frac{e_g}{e_g^2+h}\partial_a(e_g\mu^a)
 \Bigg)
 ,
 \label{mu-symmetry_finite}
\end{align}
which approaches
\begin{align}
 \delta e_g
 =\frac{1}{2}
 \left(
 \pm\sqrt{
 4e_g\partial_a(e_g\mu^a)
 +(\partial_a(e_g\mu^a))^2}
 +\partial_a(e_g\mu^a)
 \right)
 ,
\end{align}
on-shell but finite $\mu^a$, where $\pm$ is just determined by the sign of $\frac{e_g^2+h}{e_g}$ 
as $e_g^2$ approaches $-h$.
Moreover, 
by a straightforward calculation, one can prove 
$\delta e_g/e_g\geq -1$ 
for a set of finite transformations that contains identity ($\mu^a=0$).
Therefore,
this trivial local bosonic transformation does not change the sign of $e_g$.

A similar argument of the above apparent on-shell singularity 
applies to the local fermionic symmetry \eqref{kappa-symmetry_Schild}.
As $e_g^2$ approaches $-h$,
the transformation of $e_g$ behaves as
\begin{align}
 \delta e_g
 &\to
 2\sqrt{2ie_g\bigg[
 \delta\varphi^{T}\slashed\Pi_{b}
 \left( e_gh^{ab}\partial_{a}\psi-\varepsilon^{ab}\partial_a\varphi\right)
 +\delta\psi^{T}\slashed\Pi_{b}
 \left( e_gh^{ab}\partial_{a}\varphi-\varepsilon^{ab}\partial_a\psi\right)
 \bigg]}
 ,
\end{align}
for infinitesimal $\delta\varphi$ and $\delta\psi$.

Note that the local fermionic symmetry \eqref{kappa-symmetry_Schild} 
is enhanced to 32 fermionic gauge degrees of freedom 
only at the classical level.
Though it seems the fermionic degrees of freedom 
can be completely eliminated by the enhanced gauge symmetry,
it should be prohibited quantum mechanically
because the path integral equivalent to the Nambu-Goto type is not
invariant under 
a change of gauge-fixing functions for the 32 fermionic degrees of freedom
but for the 16 of them
corresponding to the fermionic first-class constraints in
Eq.~\eqref{chi_f1} and \eqref{chi_f2}.

Similarly, the path integral \eqref{partition-fn_b_PtoS} is not invariant under 
a change of gauge fixing for the trivial local bosonic transformation \eqref{mu-symmetry} as well,
and in fact, the measure \eqref{measure_eg} of $e_g$ is not invariant.
Therefore, half of the enhanced local fermionic symmetry \eqref{kappa-symmetry_Schild} and the local bosonic symmetry \eqref{mu-symmetry} are not the gauge symmetry to be maintained in the path integral we are interested in.

As a side note,
the algebra of the local fermionic symmetry \eqref{kappa-symmetry_Schild} is closed off-shell
for $\delta\psi=0$ or $\delta\varphi=0$,
together with the reparametrisation and the trivial local bosonic symmetry \eqref{mu-symmetry},
as explained in appendix \ref{sec:closure-of-gauge-trf}.
This is interesting because it has been known that
the algebra of the $\kappa$ symmetry is closed only on-shell \cite{Green:1983sg,Kallosh:1988zn,Kallosh:1989yv}
in the Nambu-Goto-type and Polyakov-type formulations.
This allows us to construct 
a covariantly formulated Schild-type theory by the BRST quantisation,
not by resorting to the Batalin-Vilkovisky formalism,
if we redefine the path integral to be invariant under the local bosonic symmetry \eqref{mu-symmetry}.

\subsection{Gauge-fixed path integral}

To obtain the gauge-fixed path integral,
let us fix the gauge symmetry 
by
\begin{align}
 e_g(\sigma)=\hat e_g(\sigma)
 ,
 \qquad
 \varphi(\sigma)=0
 ,
 \label{gauge-fixing}
\end{align}
where $\hat e_g$ is not a path-integrated field but just a function on the world-sheet.
The gauge fixing is achieved by
gauge transformations \eqref{reparam-gauge-symmetry} and \eqref{kappa-gauge-symmetry} parameterised
by $\omega(\sigma)$ and $\kappa^\varphi(\sigma)$
with $\kappa^\psi(\sigma)=0$
and the decomposition of the reparametrisation transformation
\begin{equation}
 \delta\sigma^a
 =\frac{\varepsilon^{ab}}{\hat e_g}\partial_b\xi
 +\delta\sigma_{\perp}^a
 ,
 \label{repram-decomposition}
\end{equation}
where $\frac{1}{\hat e_g}\partial_a(\hat e_g\delta\sigma_{\perp}^a)=\omega$.
It is then obvious that the Faddeev-Popov determinant\footnote{
The area-preserving diffeomorphism is left unfixed here for the Faddeev-Popov method.
This is possible if we fix the gauge by a delta functional as in Eq.~\eqref{partition-fn_gauge-fixed}.
Likewise, the whole diffeomorphism with $\omega$ can be left unfixed if $\varphi$ is fixed by the delta functional.
} is independent of $X^\mu$ and $\theta^A$.

However, there is a non-triviality in gauge-fixing $e_g$:
it cannot be fixed to only one function 
because the two regions of integration, $e_g>0$ and $e_g<0$, are
disconnected in terms of the gauge symmetry,
which can be understood also by the fact that
the integral of infinitesimal gauge transformations does not change the sign of $e_g$.
Therefore,
we fix $e_g$ by $e_g^2=\hat e_g^2$, 
with $\hat e_g(\sigma)>0$ at any world-sheet point $\sigma$.
As discussed in section~\ref{sec:path-integral},
the contributions from $e_g>0$ and from $e_g<0$ correspond to an F-string
and an anti-F-string, respectively.
The partition function is written as\footnote{
Here, we ignore integration over moduli space.
}
\begin{align}
 Z&=\mathcal{C}\int 
 \mathcal{D}X\, \mathcal{D}\theta\,
 \mathcal{D}e_g\,
 \Big[\prod_{\sigma}2\hat e_g^2\delta(e_g^2-\hat e_g^2)\delta(\varphi)\Big]
 \exp\bigg[ iS_{\rm Schild}[X,\psi,\varphi,e_g] \bigg]
 \nonumber \\
 &=\mathcal{C}\int 
 \mathcal{D}X\, \mathcal{D}\psi
 \left[ \prod_\sigma\frac{2\pi\hat e_g^{-\frac{1}{2}}}{\sqrt{\iDelta\Sigma}}
 \left(
 e^{\frac{\pi i}{4}}
 e^{ i\iDelta\Sigma\mathcal{L}_{\rm Schild}[X,\psi,0,\hat e_g]
 }
 +
 e^{-\frac{\pi i}{4}}
 e^{ i\iDelta\Sigma\mathcal{L}_{\rm Schild}[X,\psi,0,-\hat e_g]
 }
 \right)
 \right]
 \nonumber \\
 &=\mathcal{C}'\int 
 \mathcal{D}X\, \mathcal{D}\psi
 \bigg[ \prod_\sigma \sum_{s(\sigma)=\pm 1} \bigg]
 \left[ \prod_\sigma e^{s(\sigma)\frac{\pi i}{4}} \right]
 e^{ i\hat S_{\rm Schild}[X,\psi,s]}
 ,
 \label{partition-fn_gauge-fixed}
\end{align}
where $\mathcal{C}$ and $\mathcal{C'}$ are normalisation factors, 
$\mathcal{L}_{\rm Schild}$ is the Lagrangian density in the action $S_{\rm Schild}$ and
$\hat S_{\rm Schild}$ is the action gauge-fixed by $e_g=s\hat e_g$,
\begin{align}
 \hat S_{\rm Schild}[X,\psi,s]
 =\frac{1}{2\pi}\int \dif{^2\sigma} \, s\; \hat e_g\bigg[
 \frac{1}{4}\{ X^\mu,X^\nu \}_{\rm\hat P}\{ X_\mu,X_\nu \}_{\rm\hat P}
 +2is\,\psi^T\Gamma_\mu
 \{ X^\mu, \psi \}_{\rm\hat P}
 -\frac{1}{2}
 \bigg]
 ,
 \label{Schild-action_gauge-fixed}
\end{align}
using the Poisson bracket defined by
\begin{align}
 \{ f_1, f_2 \}_{\rm\hat P}
 =\frac{\varepsilon^{ab}}{\hat e_g}\partial_af_1 \partial_bf_2
 .
 \label{PB_gauge-fixed}
\end{align}
Here, we omit the cut-off regulators just for simplicity.

Note that the classical solution of $s$ is constant.
One can see from the equations of motion in the gauge $e_g=s \hat e_g$,
\begin{align}
 \partial_a\left(
 s\{ X_\mu,X_\nu\} \varepsilon^{ab}\partial_bX^\nu
 +2i\varepsilon^{ab}\psi^T\Gamma_\mu\partial_b\psi
 \right) =0,
 \quad
 \varepsilon^{ab}\Gamma_\mu\partial_aX^\mu\partial_b\psi
 =0,
\end{align}
with the constraint $\{ X^\mu,X^\nu\}^2=-1$,
that
\begin{equation*}
 0=
 \partial_c\left(
 s\{ X_\mu,X_\nu\} \right)
 \{ X^\mu,X^\nu\}
 =
 (\partial_cs)
 \{ X_\mu,X_\nu\}^2
 +\frac{1}{2}s\partial_c\left(
 \{ X_\mu,X_\nu\}^2 \right)
 ,
\end{equation*}
so that $\partial_as=0$.
Therefore,
the dominant configurations should be 
either $s(\sigma)=1$ or $s(\sigma)=-1$ at every point $\sigma$,
which is a natural conclusion.

The integral over $X^\mu$ can be decomposed into two parts:
the region with $h<0$ and that with $h>0$.
Though not obvious from the expression \eqref{partition-fn_gauge-fixed},
the integral with $h>0$ vanishes because of the cut-off introduced by the ground-state wave function,
as seen in section \ref{sec:path-integral}.
In fact, one can check 
that the configurations with $h>0$ do not contribute to the path integral 
\eqref{partition-fn_gauge-fixed}
by analytically continuing it by 
$\sigma^0=e^{-is\theta}\sigma^2$ with $\theta\to\pi/2$,
assuming that the limit as $\iDelta\Sigma\to 0$ is independent of its direction.
However, the cancellation 
for $h>0$ is not necessarily guaranteed
without the explicit cut-off regulators.
To avoid this complication, 
we explicitly insert the Heaviside step function $\theta(-h)$ into the path integral
instead of the regulators:
\begin{align}
 Z=\mathcal{C}'\int 
 \mathcal{D}X\, \mathcal{D}\psi\,
 \bigg[ \prod_\sigma \sum_{s(\sigma)=\pm 1} \bigg]
 \left[ \prod_\sigma e^{s(\sigma)\frac{\pi i}{4}} \right]
 \Big[\prod_{\sigma}\theta(-h)\Big]
 e^{ i\hat S_{\rm Schild}[X,\psi,s]}
 .
 \label{partition-fn_gauge-fixed_theta}
\end{align}
This is the partition function equivalent to the one for the Minkowskian type IIB superstring theory with the Polyakov-type action with the physically motivated cut-off regulators.

\subsection{Equivalence to the Euclidean path integral}
\label{sec:equiv-to-Euc_gauge-fixed}
In this subsection,
we show that the Minkowskian path integral 
is equivalent to its Euclidean version 
for critical bosonic string theory (without gauge-fixing)
and for critical type IIB superstring theory\footnote{
One can show the equivalence for type IIA theory
by fixing the gauge symmetry by
$\theta^2=\bar\theta^{1 \dagger}$,
so that $\Pi_a^0=\partial_aX^0$
and thus the contour deformation \eqref{Wick-rotation_S} rotates it 
in the same way as the bosonic case.
} by fixing the $\kappa$ symmetry by $\varphi=0$.
The corresponding Euclidean theory is defined by a simple Wick rotation of the original Minkowskian theory, 
without making the fermions be $Spin(10)$ spinors for the superstring case.

Let us start with a gauge-fixed Minkowskian path integral with the Polyakov-type action.
One can gauge-fix $\varphi$ in the Polyakov-type superstring theory
just like the Schild-type case,
by a $\kappa$ transformation \eqref{kappa-symmetry_P} parameterised by $\kappa^\varphi=\frac{1}{2}(\kappa^1+i\kappa^2)$ 
so that the Faddeev-Popov determinant is independent of all dynamical variables.
Hence, the gauge-fixed path integral is written as
\begin{align}
 Z&=\mathcal{C}_{\kappa}\int 
 \mathcal{D}X\, \mathcal{D}\theta\,
 \mathcal{D}g\,
 \Big[\prod_{\sigma}\delta(\varphi)\Big]
 \exp\left[ iS_{\rm P} \right]
 \nonumber \\
 &=\mathcal{C}_{\kappa}\int 
 \mathcal{D}X\, \mathcal{D}\psi\,
 Z_{\rm b}[X]\,
 \exp\bigg[
 -\frac{1}{\pi}\int \dif{^2\sigma} \, 
 \varepsilon^{ab}
 \psi^T\Gamma_\mu\partial_a X^\mu\partial_b\psi
 \bigg]
 ,
 \label{partition-fn_gauge-fixed_only-phi}
\end{align}
where
$\mathcal{C}_\kappa$ is a normalisation factor and
$Z_{\rm b}[X]$ is equal to $Z_{\rm b}[X,\theta]$ with $\varphi=0$.
Since it is not straightforward to show the equivalence to its Euclidean version
by a naive Wick rotation,
we rewrite $Z_{\rm b}[X]$ first with the Schild-type action as
\begin{align}
 Z_{\rm b}[X]
 &=
 \left[ \prod_{\sigma}
 \int_{-\infty}^{\infty}\frac{-2\pi\dif e_g}{(ie_g)^\frac{3}{2}\sqrt{\iDelta\Sigma}}
 \right]
 \exp\bigg[
 -\frac{i}{2\pi}\int \dif{^2\sigma}\,
 \frac{1}{2}\left(
 -\frac{\Sigma^{\mu\nu}\Sigma_{\mu\nu}}{2e_g}
 +e_g
 -i\epsilon'|e_g|
 -\frac{i\tilde\epsilon'}{|e_g|}
 \right)
 \bigg]
 \label{partition-fn_b_S}
 ,
\end{align}
where $\Sigma^{\mu\nu}=\varepsilon^{ab}\partial_aX^\mu\partial_bX^\nu$,
using Eq.~\eqref{partition-fn_b_PtoS}.
To relate it to the Euclidean path integral,
we deform the contour by
\begin{align}
 e_g= e^{-i\theta\sgn(e_g^{\rm (E)})}e_g^{\rm (E)},
 \qquad
 X^0= \sgn(e_g^{\rm (E)})\, e^{-i\theta\sgn(e_g^{\rm (E)})}X^{{\rm (E)}D},
 \label{Wick-rotation_S}
\end{align}
with $\theta$ rotated from 0 to $\pi/2$
and $e_g^{\rm (E)}$ running from $-\infty$ to $+\infty$.
We flip the sign of $X^0$ for negative $e_g^{\rm (E)}$ and 
rotate the fermion by $\psi=e^{i\theta/2}\psi^{\rm (E)}$,
just to make the expression of the action standard.
In addition, we relabel $\sigma^0$ as $\sigma^2$,
and use $\varepsilon^{12}=1=\varepsilon^{01}$.
Then the integrand becomes
\begin{align}
 &\exp\left[
 \frac{i}{4\pi}
 \int \dif{^2\sigma} \, \bigg(
 e^{i\theta}\frac{(\Sigma^{{\rm (E)}ij})^{2}}{2e_g^{\rm (E)}}
 -e^{-i\theta}\frac{(\Sigma^{{\rm (E)}D\,i})^{2}}{e_g^{\rm (E)}}
 -e^{-i\theta} e_g^{\rm (E)}
 +ie^{-i\theta}\epsilon'e_g^{\rm (E)}
 +ie^{i\theta}\frac{\tilde\epsilon'}{e_g^{\rm (E)}}
 \bigg)
 \right]
 ,
\end{align}
where $i,j=1,\cdots,D-1$,
for positive $e_g^{\rm (E)}$,
and thus the real part of the exponent is negative-definite\footnote{
If the $\kappa$ symmetry is not fixed,
the simple Wick rotation \eqref{Wick-rotation_S} cannot keep
$\Sigma^{{\rm (E)}D\,i}=\varepsilon^{ab}\Pi_a^{D}\Pi_b^i$ 
real
because $e^{-i\theta\sgn(e_g^{\rm (E)})}$ is not factored out of $\Pi_a^0$.
} for $0\leq\theta\leq\frac{\pi}{2}$. 
The same goes for negative $e_g^{\rm (E)}$.

By the deformation \eqref{Wick-rotation_S},
Cauchy's integral theorem equates
the path integral \eqref{partition-fn_gauge-fixed_only-phi}
to a Euclidean path integral 
as
\begin{align}
 Z
 &=\mathcal{C}_{\kappa}
 \int \mathcal{D}X^{\rm (E)}
 \mathcal{D}\psi^{\rm (E)}
 \left[ \prod_{\sigma}
 \int_0^\infty \frac{4\pi\dif e_g^{\rm (E)}}{(e_g^{\rm (E)})^\frac{3}{2}\sqrt{\iDelta\Sigma}}
 \right]
 \nonumber \\
 &\hspace{40pt}
 \times\exp\Bigg[
 -\frac{1}{2\pi}
 \int \dif{^2\sigma} \bigg(
 \frac{(\Sigma^{{\rm (E)}mn})^{2}}{4e_g^{\rm (E)}}
 +\frac{e_g^{\rm (E)}}{2}
 -2i\varepsilon^{ab}
 \psi^{{\rm (E)}T}\Gamma_\mu\partial_a X^\mu\partial_b\psi^{\rm (E)}
 \bigg)
 \Bigg]
 ,
 \label{partition-fn_b_SE}
\end{align}
where we use $\Gamma_0=i\Gamma_{D}$ and $m,n=1,\cdots,D$.
Then, 
by changing the variable $e_g^{\rm (E)}$ to $\Lambda^2=e_g^{\rm (E)}/h_{11}^{\rm (E)}$
and inserting 
\begin{align}
 1=\Det\left[ \frac{h_{11}^{\rm (E)}}{4\pi^2\Lambda^2} \right]^{\frac{1}{2}}
 \left[ \prod_\sigma
 \int_{-\infty}^{\infty}\sqrt{\iDelta\Sigma}\,\dif{\Lambda^1} \right]
 \exp\left[
 -\frac{1}{2\pi}\int\dif{^2\sigma}\frac{h_{11}^{\rm (E)}}{2\Lambda^2}\left(\Lambda^1-\frac{h_{12}^{\rm (E)}}{h_{11}^{\rm (E)}}\right)^2
 \right]
 ,
\end{align}
we obtain
\begin{align}
 Z
 &=\mathcal{C}_{\kappa}
 \int \mathcal{D}X^{\rm (E)}
 \mathcal{D}\psi^{\rm (E)}
 \left[ \prod_{\sigma}
 \frac{2\dif\Lambda^{1}\dif\Lambda^{2}}{(\Lambda^{2})^2}
 \right]
 \nonumber \\
 &\hspace{40pt}
 \times\exp\Bigg[
 -\frac{1}{2\pi}
 \int \dif{^2\sigma} \bigg(
 \frac{1}{2}\sqrt{g}g^{ab}h_{ab}^{\rm (E)}
 -2i\varepsilon^{ab}
 \psi^{{\rm (E)}T}\Gamma_\mu\partial_a X^\mu\partial_b\psi^{\rm (E)}
 \bigg)
 \Bigg]
 ,
 \label{partition-fn_b_PE}
\end{align}
where $h_{ab}^{\rm (E)}$ is defined by $h_{ab}^{\rm (E)}=\partial_aX^{{\rm (E)}m} \partial_bX_m^{\rm (E)}$ and
the Riemannian world-sheet metric is defined by
$\sqrt{g}g^{22}=\frac{1}{\Lambda^2}$ and $\sqrt{g}g^{12}=-\frac{\Lambda^1}{\Lambda^2}$.
Note that
the intervals of integration over $\Lambda^1$ and $\Lambda^2$ are 
$(-\infty,\infty)$ and $(0,\infty)$, respectively,
which guarantees that the eigenvalues of $\sqrt{g}g^{ab}$ are all positive;
$\Lambda^2$ runs over only positive values
unlike the Lorentzian case.
The measure 
$[\prod_{\sigma}\frac{\dif\Lambda^{1}\dif\Lambda^{2}}{(\Lambda^{2})^2}]$
is effectively equivalent to the standard measure of the Riemannian world-sheet metric
with the conformal factor, 
in the critical case ($D=26$ for the bosonic string and $D=10$ for the superstring).

We have shown that the Euclidean path integral \eqref{partition-fn_b_PE}
with the Polyakov-type action on the Riemannian world-sheet
is equivalent to the gauge-fixed Minkowskian path integral \eqref{partition-fn_gauge-fixed_only-phi}.
Note that the Euclidean path integrals 
\eqref{partition-fn_b_SE} and \eqref{partition-fn_b_PE}
are not exactly obtained by the gauge fixing 
because the $\kappa$ symmetry is not real anymore in the Euclidean theory.

Incidentally,
we show this is equivalent to the path integral with the Euclidean Nambu-Goto-type action as well.
By integrating out $e_g^{\rm (E)}$ in \eqref{partition-fn_b_SE} 
using \eqref{eg-int-formula},
one obtains 
\begin{align}
 Z
 &=\mathcal{C}_{\kappa}
 \int \mathcal{D}X^{\rm (E)}
 \mathcal{D}\psi^{\rm (E)}
 \left[ \prod_{\sigma}
 \frac{8\pi^2}{\iDelta\Sigma\sqrt{h^{\rm (E)}}}
 \right]
 \nonumber \\
 &\hspace{40pt}
 \times
 \exp\left[
 -\frac{1}{2\pi}
 \int \dif{^2\sigma} \left(
 \sqrt{h^{\rm (E)}}
 -2i\varepsilon^{ab}
 \psi^{{\rm (E)}T}\Gamma_\mu\partial_a X^\mu\partial_b\psi^{\rm (E)}
 \right)
 \right]
 ,
\end{align}
where 
$h^{\rm (E)}=\det h_{ab}^{\rm (E)}=\frac{1}{2}(\Sigma^{{\rm (E)}mn})^{2}$.

Let us finally gauge-fix $e_g^{\rm (E)}$ in the Euclidean Schild-type theory.
By the Euclidean version of a reparametrisation transformation,
we fix gauge symmetry in the path integral \eqref{partition-fn_b_SE}
by $e_g^{\rm (E)}=\hat e_g$
and obtain 
\begin{align}
 Z
 &=\mathcal{C}'_\kappa\int 
 \mathcal{D}X^{\rm (E)} \mathcal{D}\psi^{\rm (E)}\,
 e^{ -\hat S_{\rm Schild}^{\rm (E)}[X^{\rm (E)},\psi^{\rm (E)}]}
 ,
 \label{partition-fn_gauge-fixed_Euc}
\end{align}
where 
\begin{align}
 \hat S_{\rm Schild}^{\rm (E)}[X,\psi]
 =\frac{1}{2\pi}\int \dif{^2\sigma} \, \hat e_g
 \left[
 \frac{1}{4}(\{ X^m, X^n \}_{\rm\hat P}^{\rm (E)})^{2}
 -2i\,\psi^{T}\Gamma_m\{ X^m, \psi \}_{\rm\hat P}^{\rm (E)}
 +\frac{1}{2}
 \right] .
 \label{Schild-action_Euc}
\end{align}

\section{Path integral for the matrix model}
\label{sec:matrix-reg}
So far, we have discussed the perturbative string theory around the 10-dimensional flat spacetime.
Since we define the Minkowskian path integral 
with the appropriate cut-off regulators
in section \ref{sec:path-integral},
we can define a scattering amplitude in the Minkowski signature as
\begin{align}
 \mathcal{A}_{j_1,\cdots,j_n}(k_1,\cdots,k_n)
 =\sum_{\chi=2,0,-2,\cdots}g_s^{-\chi}
 \int \mathcal{D}X\,\mathcal{D}\theta\,\mathcal{D}e_g\,
 \mathcal{V}_{j_1}(k_1)\cdots \mathcal{V}_{j_n}(k_n)
 \exp[iS_{\rm Schild}]
 ,
\end{align}
where $g_s$ is the string coupling in the background spacetime we consider
and $\mathcal{V}_{j_i}(k_i)$ are gauge-invariant vertex operators.
The path integral for a fixed Euler characteristic $\chi$ is 
computed on a compact world-sheet of fixed topology 
with punctures corresponding to the vertex operators.

The problem is that this amplitude 
is expressed only by a perturbative expansion and that
the path integral computes the amplitude only around the flat spacetime.
To conquer this problem,
matrix models were proposed as non-perturbative formulations,
and in particular,
as a matrix-regularised version of 
the Schild-type action of fixed topology,
the IKKT matrix model was proposed.
By the matrix regularisation procedure,
all contributions of multi-string states with different topology 
are considered to be included 
as different matrix configurations generated by the matrix-model path integral \cite{Ishibashi:1996xs}.

In this section, we regularise the Schild-type theory on a spherical world-sheet
by matrices
and obtain the path integral formulation for the matrix model.
In the superstring case, 
unlike the supermembrane case for the BFSS matrix model \cite{deWit:1988wri,Banks:1996vh},
the induced metric to be regularised is Lorentzian
so that matrix regularisation is not quite straightforward.

One rigorous way to achieve matrix regularisation of the Minkowskian theory is
to deform the contour of the path integral 
by the Wick rotation \eqref{Wick-rotation_S} 
before the regularisation
so that the theory becomes Euclidean,
which is equivalent to the original Minkowskian path integral
as explained in section \ref{sec:equiv-to-Euc_gauge-fixed}.
Then one can set the world-sheet coordinates to parameterise a compact Riemann surface,
and thus the dynamical variables are matrix-regularised 
by the standard procedure.

Another way is to perform formal matrix regularisation 
keeping the Minkowski signature for the target space.
This is also a reasonable way because 
the world-sheet coordinates $\sigma^a$ are just parameters in the Schild-type theory
as well as the Nambu-Goto-type theory.
Hence, while the world-sheet has topology, 
it has neither a Lorentzian nor Riemannian metric a priori;
only one additional structure the world-sheet has in the Schild-type case 
is the ``volume-element factor''~$e_g$.
This allows us to regard $X^\mu(\sigma)$ and $\psi(\sigma)$ as 
a map from a compact Riemann surface with punctures 
to the dynamical variables.
Since we consider scattering amplitudes, there are always punctures,
and thus such a map from the Riemannian world-sheet 
to a Lorentzian target-space configuration is allowed to exist.
Therefore, applying the matrix regularisation procedure to the map without Wick rotation
is always possible for scattering amplitudes.

In the following,
we explain the applications of matrix regularisation described above:
matrix regularisation \textit{after} the Wick rotation
and that \textit{without} Wick rotation.

\subsection{Matrix regularisation after Wick rotation}
\label{sec:mat-reg_after-W}
After the Wick rotation,
the path integral becomes Eq.~\eqref{partition-fn_gauge-fixed_Euc}.
To apply the matrix regularisation procedure \cite{Hoppe82mr},
we here consider the path integral of spherical topology for simplicity.
If we set $\hat e_g$ constant, 
we describe the world-sheet as a two-sphere by
$x^1=\sqrt{\hat e_g\sigma^2(2-\hat e_g\sigma^2)}\cos\sigma^1$,
$x^2=\sqrt{\hat e_g\sigma^2(2-\hat e_g\sigma^2)}\sin\sigma^1$, and
$x^3=\hat e_g\sigma^2-1$,
which satisfy $\sum_{i=1}^3x^ix^i=1$ and 
$\{x^i,x^j\}_{\rm\hat P}^{\rm (E)}=\varepsilon_{ijk}x^k$.
We then regularise the sphere by a map,
$x^i\mapsto\frac{2}{N}L_i$ with $L_i$ 
being the $SU(2)$ generators in the $N$-dimensional irreducible representation, 
which satisfy $[L_i,L_j]=i\varepsilon_{ijk}L_k$.
For a general function on the sphere, written as 
$f(\sigma)=\sum_{l=0}^\infty\sum_{m=-l}^l f_{lm}Y_{lm}(x(\sigma))$,
the matrix regularisation is a map to a matrix
$f=\sum_{l=0}^{N-1}\sum_{m=-l}^l f_{lm}\hat Y_{lm}$,
where the matrix $\hat Y_{lm}$ is 
obtained by replacing $x^i$ 
in the spherical harmonics $Y_{lm}(x)$
with $\frac{2}{N}L_i$.
In the matrix regularisation procedure,
the multiplication of functions is mapped to the matrix multiplication,
and the Poisson bracket and the world-sheet integration are mapped via
\begin{align}
 \{ \cdot , \cdot \}_{\rm\hat P}^{\rm (E)} \mapsto
 \frac{N}{2i}[ \cdot, \cdot ],
 \qquad
 \frac{1}{4\pi}\int \dif{^2\sigma}\, \hat e_g \mapsto \frac{1}{N}\tr
 .
\end{align}

By the above procedure,
one obtains the matrix-model action
from the Euclidean Schild-type action \eqref{Schild-action_Euc} as
\begin{align}
 \hat S_{\rm Schild}^{\rm (E)}[X,\psi]
 &\mapsto 2\tr \left(
 -\frac{N}{16}[ X^m, X^n ]^2
 -\psi^{T}\Gamma_m[ X^m, \psi]
 +\frac{1}{2N}
 \right)
 \nonumber \\
 &\to -N
 \tr \left( 
 \frac{1}{4}[ X^m, X^n ]^2
 +\frac{1}{2}\psi^{T}\Gamma_m[ X^m, \psi]
 \right)
 =:S_{\rm IKKT}^{\rm (E)}[X,\psi]
 ,
 \label{IKKT-action_E}
\end{align}
with rescaling
$X^m\to 2^{\frac{1}{4}}X^m$,
$\psi\to 2^{-\frac{9}{8}}\sqrt{N} \psi$,
removing 
the constant term $\tr\frac{1}{N}$
because it does not affect the dynamics\footnote{
The constant term becomes meaningful if we define the matrix-model path integral
with summation over the matrix size $N$, which is interpreted as integration over $e_g$ in the original paper \cite{Ishibashi:1996xs}.
Since we consider, in this paper, the matrix model obtained by the matrix regularisation of the theory gauge-fixed by $e_g^{\rm (E)}=\hat e_g$, it is natural to define it without the summation.
}.

The remaining gauge symmetry after the gauge fixing \eqref{gauge-fixing}
is the area-preserving diffeomorphism, 
which is the component of $\xi$ in the transformation \eqref{repram-decomposition}.
Since the measure of the gauge-fixed path integral is invariant 
under an area-preserving transformation, 
the matrix regularisation of the measure should be invariant 
under its matrix-model counterpart, 
namely an $SU(N)$ transformation.
Therefore,
the path integral becomes
\begin{align}
 Z
 &\mapsto \int 
 \mathcal{D}X^{\rm (E)}\, \mathcal{D}\psi^{\rm (E)}\,
 e^{ -S_{\rm IKKT}^{\rm (E)}[X^{\rm (E)},\psi^{\rm (E)}]}
 ,
\end{align}
where the measure is the standard matrix-model measure, 
invariant under an $SU(N)$ transformation.
This is a well-defined finite integral for finite $N$ \cite{Krauth:1998xh,Austing:2001pk}.
One can obtain a matrix-regularised scattering amplitude 
by inserting the vertex operators
obtained by the matrix regularisation after the Wick rotation,
to the above path integral.

The Euclidean matrix-model path integral can be Wick-rotated back
to its Minkowskian version by Cauchy's integral theorem.
We deform the contour of the matrix integration by
\begin{align}
 X^{{\rm(E)}\,10}=-e^{-\frac{3 i}{4}\theta_0}X^0
 , \qquad
 X^{{\rm (E)}\, i}=e^{\frac{i}{4}\theta_1}X^{i}
 , \qquad
 \psi^{\rm (E)}=e^{\frac{3i}{8}\theta_2}\psi
 .
 \label{Wick-rotation_MM}
\end{align}
Then, the action becomes
\begin{align}
 S_{\rm IKKT}^{\rm (E)}[X^{\rm (E)},\psi^{\rm (E)}]
 &=N\tr \bigg( 
 -\frac{e^{-\frac{i}{2}(3\theta_0-\theta_1)}}{2}[ X^{0}, X^i ]^2
 -\frac{e^{i\theta_1}}{4}[ X^i, X^j ]^2
 \nonumber \\
 &\hspace{45pt}
 -i\frac{e^{\frac{3i}{4}(-\theta_0+\theta_2)}}{2}\psi^{T}\Gamma_{0}[ X^{0}, \psi]
 -i\frac{e^{\frac{i}{4}(\theta_1+3\theta_2-2\pi)}}{2}\psi^{T}\Gamma_i[ X^i, \psi]
 \bigg)
 \nonumber \\
 &
 \to
 -iN\tr \bigg( 
 -\frac{1}{2}[ X^{0}, X^i ]^2
 +\frac{1}{4}[ X^i, X^j ]^2
 \nonumber \\
 &\hspace{70pt}
 +\frac{1}{2}\psi^{T}\Gamma_{0}[ X^{0}, \psi]
 +\frac{1}{2}\psi^{T}\Gamma_i[ X^i, \psi]
 \bigg)
 =:-iS_{\rm IKKT}
 \label{IKKT-action_Wick-rot}
 ,
\end{align}
by taking $\theta_0=\theta_1=\theta_2\to\pi/2$.
Since $\tr(-[X^{0}, X^i]^2)$ and $\tr(-[X^i, X^j]^2)$ are positive semi-definite,
Cauchy's theorem equates the Euclidean path integral to its Minkowskian version
by the deformation \eqref{Wick-rotation_MM} of the contour
keeping $\theta_0$ and $\theta_1$ satisfying 
$-\pi\leq 3\theta_0-\theta_1\leq \pi$
and
$-\frac{\pi}{2}\leq \theta_1\leq \frac{\pi}{2}$,
as long as the inserted operators have no singularity that affects the deformation.
Note that the Minkowskian action \eqref{IKKT-action_Wick-rot} implicitly has natural cut-off regulators, 
$i\epsilon (X^iX^i+X^0X^0)$.

A scattering amplitude computed by insertion of $\mathcal{O}$ is then
\begin{align}
 \mathcal{A}_{\mathcal{O}}
 &=\int \mathcal{D}X^{\rm (E)}\mathcal{D}\psi^{\rm (E)}
 \mathcal{O}(-iX^{{\rm (E)}10},X^{{\rm (E)}i},e^{\frac{\pi i}{4}}\psi^{(E)})
 \, e^{-S_{\rm IKKT}^{\rm (E)}}
 \nonumber \\
 &\propto 
 \int \mathcal{D}X\,\mathcal{D}\psi\,
 \mathcal{O}(e^{\frac{\pi i}{8}}X^{0},e^{\frac{\pi i}{8}}X^i,e^{\frac{7\pi i}{16}}\psi)
 \, e^{iS_{\rm IKKT}}
 .\label{scattering-amp_IKKT_L}
\end{align}
Here, we assumed the form of the functional $\mathcal{O}$ is originated
from the Minkowskian perturbative string theory.

\subsection{Matrix regularisation without Wick rotation}
As explained,
we apply the matrix regularisation procedure
to the Schild-type theory 
on a unit two-sphere with punctures, 
keeping the Minkowski signature for the target space.
However, 
if we use the gauge-fixed Schild-type action \eqref{Schild-action_gauge-fixed},
matrix regularisation of the function of signs, $s(\sigma)$, 
is problematic.

One possible way to avert this problem
would be to fix $s(\sigma)=1$ or $s(\sigma)=-1$ at every point $\sigma$,
which is the classical configuration.
By the matrix regularisation procedure,
the gauge-fixed action \eqref{Schild-action_gauge-fixed}
becomes
the Minkowskian IKKT action $S_{\rm IKKT}$ \eqref{IKKT-action_Wick-rot}
up to a sign of the bosonic term.
Since the contributions from both $s=1$ and $-1$ 
should enter the path integral \eqref{partition-fn_gauge-fixed_theta},
the matrix-regularised path integral is
\begin{align}
 &\int 
 \mathcal{D}X\, \mathcal{D}\psi\;
 \Theta\big( [ X^\mu,X^\nu ]^2 \big)
 \cos\left[
 N\tr\left(
 \frac{1}{4}[ X^\mu,X^\nu ]^2
 \right)
 +\text{const.}
 \right]
 \nonumber \\
 &\hspace{160pt}
 \times
 \exp\left[
 iN\tr\left(
 \frac{1}{2}\,\psi^T\Gamma_\mu
 [ X^\mu, \psi ]
 \right)
 \right]
 ,
\end{align}
where $\Theta$ is a function that takes 1 if all the eigenvalues of the matrix-valued argument are not negative and 0 otherwise.
However, this does not fully take account of the quantum dynamics of $s(\sigma)$,
namely the contributions from both an F-string and an anti-F-string.

To resolve the problem of $s(\sigma)$ completely,
we matrix-regularise the Schild-type action 
before the gauge fixing of $e_g$, 
which leads us to the NBI matrix model \cite{Fayyazuddin:1997yf} in the Minkowski signature.
The path integral \eqref{partition-fn_gauge-fixed}
without the gauge fixing $[2e_g^2\delta(e_g^2-\hat e_g^2)]$
(or equivalently, the path integral \eqref{partition-fn_gauge-fixed_only-phi}
with Eq.~\eqref{partition-fn_b_S})
is matrix-regularised as
\begin{align}
 Z&\mapsto
 \int 
 \mathcal{D}X\, \mathcal{D}\psi\,
 \mathcal{D}Y\,
 e^{
 iS_{\rm NBI}[X,\psi,Y]
 }
 ,
 \label{partition-fn_NBI}
\end{align}
where $\mathcal{D}X\,\mathcal{D}\psi\,\mathcal{D}Y$ is the standard $SU(N)$-invariant measure,
by matrix-regularising $e_g\mapsto -Y$.
Here, the action of the ``dielectric'' NBI matrix model is
\begin{align}
 S_{\rm NBI}[X,\psi,Y]
 &=N\tr \bigg( 
 \frac{1}{4}Y^{-1}[ X^\mu, X^\nu ]^2
 +\frac{1}{2}\psi^{T}\Gamma_\mu[ X^\mu, \psi]
 +\frac{1}{N^2}Y
 \nonumber \\
 &\hspace{120pt}
 +\frac{i}{N}\left( N+\frac{1}{2} \right) \ln (-iY)
 +i\epsilon Y^2
 +i\tilde\epsilon Y^{-2}
 \bigg)
 ,
 \label{NBI-MMaction}
\end{align}
where we explicitly introduce the cut-off terms with $\epsilon$ and $\tilde\epsilon$.
Unlike the original NBI matrix model,
we take $Y$ as a Hermitian matrix that is not positive-definite but can have negative eigenvalues,
which reflects the original interval of integration over $e_g$.

In the matrix regularisation process,
we assume that the measure of $e_g$ provides
the term $i\ln(-iY)$ with its coefficient being $N+\frac{1}{2}$ 
so that 
the path integral \eqref{partition-fn_NBI}
does not contain 
configurations in which $M:=[X^\mu,X^\nu]^2$ has one or more negative eigenvalues.
In fact, 
the integration over $Y$ is rewritten 
by a similar calculation to Ref.~\cite{Fayyazuddin:1997yf} 
as
\begin{align}
 &\int 
 \mathcal{D}Y\,
 \exp\left[
 iN\tr \bigg( 
 \frac{1}{4}Y^{-1} M
 +\frac{1}{N^2}Y
 +\frac{i}{N}\left( N+\frac{1}{2} \right) \ln (-iY)
 +i\epsilon Y^2
 +i\tilde\epsilon Y^{-2}
 \bigg)
 \right]
 \nonumber \\
 &\propto
 \Delta(m)^{-1}\det_{i,j}
 \left[
 \left( iN\pder{}{\alpha}\right) ^{j-1}
 \left(
 \frac{e^{-i\sqrt{m_i-i\epsilon'}\sqrt{\alpha}}}{\sqrt{m_i-i\epsilon'}}
 +\frac{e^{i\sqrt{m_i+i\epsilon'}\sqrt{\alpha}}}{\sqrt{m_i+i\epsilon'}}
 \right)
 \right]_{\alpha\to 1}
 ,
\end{align}
for infinitesimal $\epsilon$, $\tilde\epsilon$ and $\epsilon'$,
where $m_i$'s are the eigenvalues of $M$ and 
$\Delta(m)=\prod_{i,j<i}(m_i-m_j)$ is the Vandermonde determinant.
It is now obvious that the integral is zero 
if at least one of the eigenvalues $m_i$ is negative,
by cancellation similar to the one in perturbative string theory,
seen in Eq.~\eqref{partition-fn_b_NG}.

\section{Summary and discussion}
\label{sec:summary}
Our careful study of the path integration for perturbative string theory
showed the quantum mechanical equivalence between the Nambu-Goto-type, Schild-type and Polyakov-type formulations.
The equivalence was proven for type II superstring theory and bosonic string theory on the flat Minkowski spacetime in the critical dimension.
Remarkably, full integration over the world-sheet metric 
results in realisation of causality, 
which has a natural interpretation by 
a ``positive-energy'' F-string and a ``negative-energy'' anti-F-string.
We also showed the equivalence between the Minkowskian path integral and its Euclidean version, for type II superstring theory and bosonic string theory.
These equivalences justify the connection between the Minkowskian path integral with the Nambu-Goto-type action and 
Polyakov's Euclidean path integral.

In the realisation of causality, 
the ``negative-energy'' anti-F-string plays a crucial role.
Such an anti-F-string seems to travel backward in time 
because of the flip of the overall sign of the Nambu-Goto-type Lagrangian by $s=-1$, 
as in Eq.~\eqref{partition-fn_b_NG}.
Its possible interpretation is that
we regard it as just equivalent to the positive-energy F-string
because an anti-particle that virtually travels backward in time is 
actually a particle that travels forward in quantum field theory.
There is also a string-specific viewpoint of this issue: 
the flip of the overall sign in the Polyakov-type Lagrangian, 
namely the flip of the sign of $\Lambda^0$,
corresponds to an interchange of $\sigma^0$ and $\sigma^1$.
Hence, a ``negative-energy'' anti-F-string can be
converted to a positive-energy F-string 
by the interchange of the parameters
for a compact world-sheet.
Another implication of the causality is that,
in the Minkowskian path integral,
the target-space configuration should have singularity
at a string interaction by the splitting--joining process.
However, this will not be a problem 
because the equivalent Euclidean path integral provides
the desired features of a smooth string interaction.

We then revisited the derivation of the IKKT matrix model 
based on our novel knowledge on the path integration for perturbative string theory.
We found that 
the matrix regularisation of the gauge-fixed Schild-type theory 
\textit{after} the Wick rotation 
gives the path integral for the Euclidean IKKT matrix model.
This is equivalent to the Minkowskian IKKT matrix model with natural cut-off regulators.
In contrast,
the matrix regularisation of the Schild-type theory
\textit{without} Wick rotation
gives the path integral for the Minkowskian NBI matrix model,
with the Hermitian matrix $Y$ being not positive- or negative-definite.

A caveat is that
the matrix models derived by the matrix regularisation
are merely candidates for the non-perturbative formulation
of superstring theory that reproduces our universe.
Notice also that
the derived Minkowskian path integrals (Eq.~\eqref{scattering-amp_IKKT_L} and \eqref{partition-fn_NBI})
are implicitly regularised by cut-off terms like $\epsilon\tr X^iX^i$
since the path integral of perturbative string theory has such terms.
However, it is equally possible that
the correct non-perturbative formulation is defined 
by such a Minkowskian path integral but regularised by a different scheme---one 
recently proposed scheme is the Minkowskian matrix model whose Lorentz symmetry is ``gauge-fixed'' \cite{Asano:2024def},
the guiding principle of which is that 
the Lorentz symmetry is conserved at the quantum level.

One of the most important tasks in the matrix models
is to reproduce perturbative string theory in a complete manner.
To achieve this,
one needs to compute the matrix model to all orders.
In particular, type IIB perturbative superstring theory on the 10-dimensional flat spacetime
should be reproduced around the trivial classical solution $X^\mu=0$ of the IKKT matrix model
because it preserves $SO(9,1)$ Lorentz symmetry.
Since physics around the trivial solution will be
unaffected by the regularisation scheme of the Minkowskian path integral
according to Ref.~\cite{Asano:2024def},
one can choose the natural but Lorentz-symmetry-breaking cut-off regulators 
for the Minkowskian path integrals just to reproduce perturbative string theory.
This is expected to be computable for some sectors,
for instance by the supersymmetric localisation technique \cite{Pestun:2007rz},
though yet to be computed.

For an explicit computation for reproducing perturbative string theory,
one should identify 
the matrix version of the vertex operators in perturbative string theory,
which is naturally assumed to be written by Wilson loops \cite{Fukuma:1997en,Hamada:1997dt}.
In fact, the so-called ``dynamical'' supercharge operator in the IKKT matrix model\footnote{
See Ref.~\cite{Dasgupta:2000df} for the BFSS case.
}
forms a massless multiplet of type IIB supergravity \cite{Kitazawa:2002vh,Iso:2004nn,Kitazawa:2006rw} 
by acting onto the Wilson loop $\mathcal{V}^{\Phi}(k)=\tr\exp[ik_\mu X^\mu]$,
which corresponds to the dilaton state.
It is important to confirm, 
through the explicit computation of scattering amplitudes, that 
a perturbative string state
is indeed described by
the supersymmetric Wilson loop
and find out its exact relationship to the matrix degrees of freedom.

\section*{Acknowledgment}
The author thanks Nobuyuki Ishibashi, Goro Ishiki, Hikaru Kawai, Tsunehide Kuroki, Takaki Matsumoto, Jun Nishimura and Asato Tsuchiya for helpful discussion.
This work was supported by JSPS KAKENHI Grant Number JP24K07036.

\appendix
\section{Closure of the local transformations}
\label{sec:closure-of-gauge-trf}
In this appendix,
we show that there is a subalgebra of the local symmetry of the Schild-type action
that is closed off-shell.

Let us first define operators of the local transformations \eqref{reparam-gauge-symmetry}, \eqref{kappa-symmetry_Schild} and \eqref{mu-symmetry}:
\begin{align}
 &q^{\rm b}_v
 =\int \dif{^2\sigma} \left(
 -iv^a(\sigma)\Pi^\mu_aP_\mu
 -v^a(\sigma)\partial_a\theta^{AT}\mathcal{D}^A
 -\partial_a(e_g v^a(\sigma))W
 \right)
 ,
 \nonumber \\
 &q^{\rm f}_{\kappa}
 =\int \dif{^2\sigma} \left(
 \delta_\kappa\theta^{AT}\mathcal{D}^A
 +\delta_\kappa e_gW
 \right)
 =q^{\rm f\varphi}_{\kappa}+q^{\rm f\psi}_{\kappa}
 ,
 \nonumber \\
 &q^{\rm f\varphi}_{\kappa}
 =\int \dif{^2\sigma} \left(
 \delta_\kappa\varphi^T(\mathcal{D}^1-i\mathcal{D}^2)
 +\frac{8ie_g^2}{e_g^2+h}\delta_{\kappa}\varphi^T\slashed\Pi_b
 \left( \frac{-h}{e_g}h^{ab}\partial_a\psi
 -\varepsilon^{ab}\partial_a\varphi \right) W
 \right)
 ,
 \nonumber \\
 &q^{\rm f\psi}_{\kappa}
 =\int \dif{^2\sigma} \left(
 \delta_\kappa\psi^T(\mathcal{D}^1+i\mathcal{D}^2)
 +\frac{8ie_g^2}{e_g^2+h}\delta_{\kappa}\psi^T\slashed\Pi_b
 \left( \frac{-h}{e_g}h^{ab}\partial_a\varphi
 -\varepsilon^{ab}\partial_a\psi \right) W
 \right)
 ,
 \nonumber \\
 &q^{\rm g}_{\mu}
 =\int \dif{^2\sigma} \left(
 \frac{e_g^2}{e_g^2+h}\partial_a(e_g\mu^a(\sigma)) W
 \right)
 ,
\end{align}
where
\begin{align}
 P_\mu(\sigma)=-i\dder{}{X^\mu(\sigma)},
 \quad
 \mathcal{D}^A(\sigma)=\dder{}{\theta^A(\sigma)}-i\Gamma^\mu\theta^A(\sigma)\dder{}{X^\mu(\sigma)},
 \quad
 W(\sigma)=\dder{}{e_g(\sigma)}.
\end{align}
The (anti-)commutators between $P_\mu$, $\mathcal{D}^A$ and $W$ are
zero except for
\begin{align}
 [\mathcal{D}^A(\sigma),\mathcal{D}^B(\sigma')]_+=2\delta^{AB}\Gamma^\mu P_\mu(\sigma)\delta^2(\sigma-\sigma'),
\end{align}
where $[\ast,\ast]_+$ denotes the anti-commutator.
$q^{\rm f\varphi}_{\kappa}$ and $q^{\rm f\psi}_{\kappa}$ 
are the local fermionic symmetries restricted by 
$\delta_\kappa\psi=0$
and 
$\delta_\kappa\varphi=0$,
respectively.
One can check that
the standard $\kappa$ symmetry algebra with $\delta_{\kappa}\theta^A=(\mathbf{1}+(-)^{A+1}\frac{\varepsilon^{ab}}{2e_g}\slashed\Pi_{ab})\kappa^A$ is closed up to equations of motion:
\begin{align}
 [q^{\rm f}_{\kappa_1},q^{\rm f}_{\kappa_2}]
 &=q^{\rm b}_{v_0}+q^{\rm f}_{\kappa_0}
 -2\frac{e_g^2+h}{e_g^2}\kappa_1^{AT}\Gamma^\mu\kappa_2^AP_\mu
 \nonumber \\
 &\qquad
 +4i\partial_a\left[
 \frac{-h}{e_g}h^{ab}\Pi_{b}^{\mu}
 -2i\varepsilon^{ab}(\psi^T\Gamma^\mu\partial_b\psi
 +\varphi^T\Gamma^\mu\partial_b\varphi)
 \right] \kappa_1^{AT}\Gamma_\mu\kappa_2^A W
 ,
\end{align}
where
\begin{align}
 &v_0^a=-\frac{4i}{e_g}\left(
 \frac{-h}{e_g}h^{ab}+(-)^{A+1}\varepsilon^{ab}
 \right) \kappa_1^{AT}\slashed\Pi_b\kappa_2^A,
 \nonumber \\
 &\delta_{\kappa_0}\theta^A
 =(\delta_{\kappa_1}\delta_{\kappa_2}-\delta_{\kappa_2}\delta_{\kappa_1})\theta^A
 +v_0^a\partial_a\theta^A
 .
\end{align}
Note that 
$\delta_{\kappa_0}\theta^A$ is a summation of a term proportional to
$(\mathbf{1}+(-)^{A+1}\frac{\varepsilon^{ab}}{2e_g}\slashed\Pi_{ab})$
and a so-called $\lambda$ transformation,
up to equations of motion.
In the Schild-type action, the $\lambda$ symmetry 
is $q^{\rm f}_{\lambda}$ with
$\delta_{\lambda}\theta^A=(\frac{-h}{e_g}h^{ab}-(-)^{A+1}\varepsilon^{ab})\partial_a\theta^A\lambda_b$,
which transforms the ``volume-element factor''~$e_g$ as well,
unlike the $\lambda$ symmetry in the Polyakov-type action.

Then,
one obtains
the following closed algebra without any terms proportional to the equations of motion:
\begin{align}
 &[q^{\rm b}_{v_1},q^{\rm b}_{v_2}]
 =q^{\rm b}_{v_1^a\partial_av_2-v_2^a\partial_av_1
 +\delta_{v_1}v_2-\delta_{v_2}v_1},
 \qquad
 [q^{\rm b}_v,q^{\rm f \varphi}_\kappa]
 =q^{\rm f \varphi}_{\kappa'}
 -q^{\rm b}_{\delta_{\kappa}v},
 \qquad
 [q^{\rm f \varphi}_{\kappa_1},q^{\rm f \varphi}_{\kappa_2}]
 =q^{\rm f \varphi}_{\kappa_3}+q^{\rm g}_{\mu_3},
 \nonumber \\
 &[q^{\rm b}_{v},q^{\rm g}_{\mu}]=q^{\rm g}_{\mu'}-q^{\rm b}_{\delta_\mu v},
 \qquad
 [q^{\rm f \varphi}_{\kappa},q^{\rm g}_{\mu}]=q^{\rm f \varphi}_{\kappa''}+q^{\rm g}_{\mu''},
 \qquad
 [q^{\rm g}_{\mu_1},q^{\rm g}_{\mu_2}]=q^{\rm g}_{\mu_3'},
\end{align}
where
\begin{align}
 &\delta_{\kappa'}\varphi=\delta_{v}\delta_{\kappa}\varphi+v^a\partial_a\delta_{\kappa}\varphi
 ,
 \quad
 \delta_{\kappa_3}\varphi=(\delta^\varphi_{\kappa_1}\delta_{\kappa_2}-\delta^\varphi_{\kappa_2}\delta_{\kappa_1})\varphi,
 \quad
 \mu_3^a=-\frac{8i}{e_g}\varepsilon^{ab}\delta_{\kappa_2}\varphi^T\slashed\Pi_b\delta_{\kappa_1}\varphi,
 \nonumber \\
 &\mu'^a=\frac{1}{e_g}\partial_b(e_g\mu^b)v^a-\frac{1}{e_g}\partial_b(e_gv^b)\mu^a
 +\delta_v\mu^a,
 \nonumber \\
 &\mu''^a=\frac{8ie_g}{e_g^2+h}\delta_{\kappa}\varphi^T\slashed\Pi_c
 \left( \frac{-h}{e_g}h^{bc}\partial_b\psi
 -\varepsilon^{bc}\partial_b\varphi \right)
 \mu^a
 +\delta_\kappa \mu^a,
 \quad
 \delta_{\kappa''}\varphi=-\delta_{\mu}\delta_{\kappa}\varphi,
 \nonumber \\
 &\mu_3'^a=\frac{e_g}{e_g^2+h}\left(
 \partial_b(e_g\mu_1^b)\mu_2^a-\partial_b(e_g\mu_2^b)\mu_1^a
 \right)
 +\delta_{\mu_1}\mu_2^a-\delta_{\mu_2}\mu_1^a
 .
\end{align}
Here, we let  
the transformation parameters $v$, $\kappa$, $\mu$ 
etc.~dependent on fields in general,
which leaves $\delta_\kappa\mu^a$, $\delta_v\delta_\kappa\varphi$, 
etc.~in the expressions.
One obtains the same closure of algebra 
if we replace $q^{\rm f \varphi}_\kappa$ with $q^{\rm f \psi}_\kappa$.

To simplify the local symmetry algebra,
let us set $\delta_{\kappa}\varphi$ independent of any fields,
keep only the area-preserving diffeomorphism 
$q^{\rm a.p.}_\xi:=q^{\rm b}_{v}$ with 
$v^a=\frac{\varepsilon^{ab}}{\hat e_g}\partial_b\xi$,
and take $\mu^a=\bar\mu^a/e_g$,
where $\hat e_g$ is just a function on the world-sheet.
Then the algebra becomes closed off-shell again:
\begin{align}
 &[q^{\rm a.p.}_{\xi_1},q^{\rm a.p.}_{\xi_2}]
 =q^{\rm a.p.}_{-\{\xi_1,\xi_2\}_{\rm\hat P}},
 \qquad
 [q^{\rm a.p.}_\xi,q^{\rm f \varphi}_\kappa]
 =q^{\rm f \varphi}_{\kappa'},
 \qquad
 [q^{\rm f \varphi}_{\kappa_1},q^{\rm f \varphi}_{\kappa_2}]
 =q^{\rm g}_{\bar\mu_3/e_g},
 \nonumber \\
 &[q^{\rm a.p.}_{\xi},q^{\rm g}_{\bar\mu/e_g}]=q^{\rm g}_{\bar\mu'/e_g},
 \qquad
 [q^{\rm f \varphi}_{\kappa},q^{\rm g}_{\bar\mu/e_g}]=0,
 \qquad
 [q^{\rm g}_{\bar\mu_1/e_g},q^{\rm g}_{\bar\mu_2/e_g}]=0,
\end{align}
where $\delta_{\kappa'}\varphi=\{\delta_{\kappa}\varphi,\xi\}_{\rm\hat P}$,
$\bar\mu_3^a=-8i\varepsilon^{ab}\delta_{\kappa_2}\varphi^T\slashed\Pi_b\delta_{\kappa_1}\varphi$
and $\bar\mu'^a=\frac{\varepsilon^{ac}}{\hat e_g}\partial_c\xi\partial_b\bar\mu^b$.
$\{\ast ,\ast \}_{\rm\hat P}$ denotes the Poisson bracket \eqref{PB_gauge-fixed}.

If we consider a Schild-type theory in which the $\kappa$ symmetry by $q^{\rm f \varphi}_\kappa$ and the trivial bosonic symmetry by $q^{\rm g}_\mu$ are actual gauge symmetry together with the reparametrisation symmetry by $q^{\rm b}_v$,
we can construct a BRST-invariant theory.
The BRST transformation with a Grassmann-odd infinitesimal parameter $\epsilon$ is
\begin{align}
 &\delta X^\mu=\epsilon(\{c,X^\mu\}_{\rm\hat P}-2i\gamma^T \Gamma^\mu\psi),
 \qquad
 \delta \psi=\epsilon \{c,\psi\}_{\rm\hat P},
 \qquad
 \delta\varphi= \epsilon( \{c,\varphi\}_{\rm\hat P}+\gamma),
 \nonumber \\
 &\delta e_g=
 \epsilon \hat e_g\left\{c,\frac{e_g}{\hat e_g}\right\}_{\rm\hat P}
 +\epsilon\frac{8ie_g^2}{e_g^2+h}
 \gamma^T\slashed\Pi_b
 \left( \frac{-h}{e_g}h^{ab}\partial_a\psi
 -\varepsilon^{ab}\partial_a\varphi \right)
 +\epsilon\frac{e_g^2}{e_g^2+h}\partial_ac_g^a
 ,
 \nonumber \\
 &\delta c=\frac{\epsilon}{2} \{c,c\}_{\rm\hat P},
 \qquad
 \delta \gamma=\epsilon \{c,\gamma\}_{\rm\hat P},
 \qquad
 \delta c_g^a
 =-\frac{\epsilon}{\hat e_g}\varepsilon^{ac}\partial_cc\,\partial_bc_g^b
 +4i\epsilon\varepsilon^{ab}\gamma^T\slashed\Pi_b \gamma
 ,
 \nonumber \\
 &\delta b=\epsilon B
 ,
 \qquad
 \delta \beta=\epsilon F
 ,
 \qquad
 \delta b_g=\epsilon B_g
 ,
 \qquad
 \delta B=0
 ,
 \qquad
 \delta F=0
 ,
 \qquad
 \delta B_g=0
 ,
\end{align}
where $c$, $\gamma$ and $c_g^a$ are ghosts 
and $b$, $\beta$ and $b_g$ are anti-ghosts.
The number of ghosts matches the gauge degrees of freedom 
by counting $c_g^a$ as one 
since it appears only in the form of $\partial_ac_g^a$.
One can then define a path integral with a BRST-invariant action 
by choosing a set of gauge-fixing conditions.
Gauge-invariant vertex operators are constructed by $X^\mu+2i\varphi^T\Gamma^\mu\psi$, such as
\begin{align}
 \int \dif{^2\sigma}\, \hat e_g\, e^{ik_\mu (X^\mu+2i\varphi^T\Gamma^\mu\psi)}
 .
\end{align}

\bibliographystyle{JHEPnote}
\bibliography{ikktq}

\end{document}